\documentclass[aps,prb,groupedaddress,showpacs,twocolumn]{revtex4}

\usepackage{graphicx}
\usepackage{amsmath}
\usepackage{bbm}
\usepackage{xspace}
\usepackage{color}
\usepackage{sidecap}
\usepackage[normalem]{ulem}

\DeclareMathOperator{\Tr}{\mbox{Tr}}
\DeclareMathOperator{\tr}{\mbox{tr}}
\DeclareMathOperator{\im}{\mbox{Im}}

\definecolor{orange}{RGB}{252,77,6}
\definecolor{darkgreen}{RGB}{00,200,00}
\definecolor{brown}{RGB}{200,127,50}
\newcommand{\noeach}[1]{{ #1}}
\newcommand{\nomkch}[1]{{ #1}}

\newcommand{\pcluster}{{\cal P}}

\newcommand{\ie}{i.\thinspace{}e.\@\xspace}
\newcommand{\eg}{e.\thinspace{}g.\@\xspace}

\newcommand{\ve}[1]{{\bf #1}}

\newcommand{\nag}{{\phantom{\dagger}}}

\newcommand{\eq}[1]{Eq.\thinspace{}(\ref{#1})}

\newcommand{\fig}[1]{Fig.\thinspace{}\ref{#1}}

\newcommand{\fc}[1]{({#1})}

\DeclareMathOperator{\Gop}{G}

\newcommand{\Ga}[2]{\;{\sideset{^{}}{_{#1}^{A#2}}\Gop}\;}
\newcommand{\Gr}[2]{\;{\sideset{^{}}{_{#1}^{R#2}}\Gop}\;}

\newcommand{\re}{\text{Re}\,}

\newcommand{\vvr}{\ve r}

\newcommand{\beq}{\begin{equation}}
\newcommand{\eeq}{\end{equation}}
\newcommand{\beqn}{\begin{eqnarray}}
\newcommand{\eeqn}{\end{eqnarray}}

\usepackage{hyperref}

\begin{document}

\title{Nonequilibrium steady state for strongly-correlated many-body systems:\\
 variational cluster approach}

\author{Michael Knap}
\email[]{michael.knap@tugraz.at}
\affiliation{Institute of Theoretical and Computational Physics, Graz University of Technology, 8010 Graz, Austria}
\author{Wolfgang von der Linden}
\affiliation{Institute of Theoretical and Computational Physics, Graz University of Technology, 8010 Graz, Austria}
\author{Enrico Arrigoni}
\affiliation{Institute of Theoretical and Computational Physics, Graz University of Technology, 8010 Graz, Austria}

\date{\today}

\begin{abstract}
A numerical approach is presented that allows to compute
nonequilibrium steady state properties of strongly correlated quantum many-body systems. 
The method is imbedded in the Keldysh Green's function formalism and
is based upon the idea of the variational cluster approach as far as
the treatment of strong correlations is concerned. It appears that 
the variational aspect is crucial as it allows 
for a suitable optimization of a ``reference'' system
to the nonequilibrium target state. The approach 
is neither perturbative in the many-body interaction nor in the
field, that drives the system out of equilibrium, and it allows to
study strong perturbations and nonlinear responses
of systems in which also the correlated region is spatially extended.
We apply the presented approach to non-linear transport 
across a strongly correlated quantum wire
described by the fermionic Hubbard model.
We illustrate how the method bridges  to cluster dynamical mean-field theory 
upon coupling two baths containing and increasing number of uncorrelated sites.
\end{abstract}


\maketitle

\section{Introduction}
\label{introduction}
The theoretical understanding of the nonequilibrium behavior  of
strongly correlated quantum  many-body systems 
is a long standing
challenge, which has become  increasingly relevant with the progress
made in the fields of
quantum optics and quantum simulation,
semiconductor, quantum, and magnetic heterostructures, nanotechnology, or
spintronics.
In the field of quantum optics and quantum simulation recent advances in experiments with ultracold gases in optical
lattices shed new light on strongly-correlated many body systems and their nonequilibrium properties. In
these experiments, specific lattice Hamiltonians can be engineered and
studied with a remarkable high level of control, making strong
correlations observable on a macroscopic scale.\cite{jaksch_cold_1998,greiner_quantum_2002,bloch_many-body_2008}
In this field another very promising experimental setup to study correlation effects
are coupled cavity quantum electrodynamic systems which contain some form of optical
nonlinearity 
resulting from the interaction of light with atomic levels.\cite{hartmann_quantum_2008,to.fa.10}
These coupled cavity systems are inherently out of
equilibrium, since they are driven by external lasers and susceptible
to dissipation. 
Semiconductor, quantum, and magnetic heterostructures subject to a bias voltage also display nonequilibrium physics, where 
strong correlations
play a decisive role.
 Experiments which study   
transport in molecular junctions 
demonstrate that many-body effects, also in combination with
vibrational modes
 are crucial, 
see, \eg, Refs.~\onlinecite{pa.pa.00,pa.fl.05}.
Another class of 
material structures with remarkable
nonequilibrium properties are 
(multi-well) heterostructures of diluted magnetic semiconductors (DMSs)
and superlattices 
embedded 
in normal metals.
These systems are of great interest as they open the possibility to
tailor electronic and spintronic devices for computing and
communications based on their unique 
interplay of spin and electronic degrees of freedom. 
Moreover, they
display a 
pronounced nonlinear transport behavior.\cite{zu.fa.04,fa.ma.07,sl.go.03,bonilla_non-linear_2005,jungwirth_theory_2006,ertler_proposal_2010,ertler_self-consistent}
The source of
nonlinearity is 
also related to
the  strong interaction between charge carriers, excitations and vibrational modes.
In addition, spin degrees of freedom clearly play a major role in their
transport properties. In order to fabricate technologically useful structures the
theoretical understanding of these highly correlated quantum many-body
systems is  
indispensable. 

A typical nonequilibrium situation in all these  systems is
conveniently  described theoretically
by switching on a
perturbation at a certain time $\tau=\tau_0$, 
for example, a bias voltage, which is 
then kept constant  after a short switching time. 
For this problem 
one may, on the one hand be interested in
transient properties at short times after switching on the
perturbation, for example  in ultrafast pump-probe
spectroscopy.~\cite{so.bl.03}
In this case, 
the properties of the system depend on the initial state, as well as
on the line shape of the 
switch-on pulse.  
For longer times away from 
$\tau_0$, quite generally one expects the system to reach a
steady state, whose properties do not depend on details of the initial
state. Nonequilibrium steady states are relevant, for example, in quantum
electronic transport across heterostructures, quantum dots, molecules (see,
\eg, Refs.~\onlinecite{ha.ja,ra.sm.86,me.wi.92,me.wi.93,ry.gu.09,scho.09}) or in driven-dissipative ultracold atomic systems.\cite{di.mi.08,kr.bu.08,di.to.10,pi.da.10,tomadin_nonequilibrium_2011,barmettler_dynamical_2010}
Intriguingly, it was shown in Ref.~\onlinecite{dalla_torre_quantum_2010} that nonequilibrium noise, which is present for instance in Josephson junctions, trapped ultracold polar molecules or trapped ions, still preserves the critical nonequilibrium steady states thus being a marginal perturbation as opposed to the temperature.
Among the methods to treat strongly correlated systems out of equilibrium, one
should mention density-matrix renormalization group and related
matrix-product state methods,~\cite{wh.fe.04,daley_time-dependent_2004,prosen_matrix_2009,be.ca.09,pe.ve.07}
continuum-time quantum Monte-Carlo,~\cite{we.ok.09}
different numerical and semi-analytical renormalization-group
approaches,~\cite{an.sc.06,ja.me.07,scho.09} equation-of-motion
methods,~\cite{ha.ja,me.wi.93},  dynamical mean-field
theory,~\cite{fr.tu.06,jo.fr.08,eckstein_thermalization_2009,ar.ko.11u} scattering Bethe Ansatz,~\cite{me.an.06,gritsev_exact_2010} and the dual-fermion approach.\cite{jung_dual-fermion_2010}
Recently, Balzer and Potthoff~\cite{balzer_nonequilibrium_2011} have presented a generalization of
cluster-perturbation theory (CPT) to the Keldysh contour, which allows
for the treatment of time-dependent phenomena.
Their results show that CPT describes quite accurately the short and
medium-time dynamics of a Hubbard chain.
A detailed study of the short-time dynamics of weakly correlated electrons in quantum transport based on the time evolution of the nonequilibrium Kadanoff-Baym equations,  
where correlations are treated in Hartree-Fock-, second Born-, and GW-approximation has been given
in Ref.~\onlinecite{my.st.09}. These approximations are 
restricted to moderate correlations but  on the other hand they allow to study rather complex models and geometries.
As far as the 
steady-state  
behavior  is concerned,
the nonequilibrium (Keldysh) Green's function approach has been 
widely used on an ab-initio or tight-binding level, where correlations are treated in mean-field approximation. Since the effective particles are non-interacting,
the Meir-Wingreen expression \cite{me.wi.92} for the current can be applied, which relates the current to the retarded Green's functions of the 
scattering with a self-energy that is renormalized due to the presence of the leads. Representative applications for nano-structured materials and molecular 
devices are given in Refs.~\onlinecite{Brandbyge:2002,fu.br.08,ma.ja.09} and in the review article Ref.~\onlinecite{ry.gu.09}.

Here we aim at strongly correlated many-body systems, and we propose
a variational cluster method, that allows to study steady-state properties.

The paper is organized as follows: In Sec.~\ref{method} we  
present the variational cluster method to treat correlated systems out
of equilibrium.
After an introductory discussion as well as relation to previous work, 
we present the general method in
Sec.~\ref{vcanss}. We discuss the self-consistency
condition in Sec.~\ref{sc}. In Sec.~\ref{model} we introduce
two specific models describing a strongly correlated Hubbard chain and a strongly correlated Hubbard ladder, respectively, which are
embedded between left and right uncorrelated reservoirs with different
chemical potentials and on-site energies. 
This results in a voltage bias which is applied to the system.
Results for the steady-state current density are discussed in 
Sec.~\ref{results}. Finally, in Sec.~\ref{conclusions} we present
our conclusions and outlook.

\section{Method}
\label{method}

In order to study nonequilibrium properties of strongly correlated
systems 
one typically considers a model 
consisting of two leads with uncorrelated particles, 
and a central correlated region. The three regions are
initially decoupled. At a certain time $\tau_0$ a coupling $V$ between
the three regions is switched on.
A natural approach is to treat $V$ via strong-coupling perturbation
theory, which at the lowest order essentially corresponds to
cluster-perturbation theory (CPT).
In Ref.~\onlinecite{balzer_nonequilibrium_2011} it has been shown that the short time behavior
can be well described 
within CPT. This can be understood 
from the observation that switching  
on the inter-cluster hopping
$V$ for a certain time $\Delta \tau$ produces a perturbation of order $V\,
\Delta \tau$, which is accounted for at first order in CPT. Therefore, we expect
the result to be accurate for small $\Delta \tau$. 
When addressing the steady state it is, thus, essential to improve the
long-time behavior.
Here, we suggest that nonequilibrium CPT 
can be systematically improved by minimizing 
some suitable
``difference'' between the unperturbed (``reference'') state which enters CPT and
the target steady state.

The strategy presented here to achieve this goal 
consists in exploiting the fact that the 
decomposition of the
Hamiltonian into an ``unperturbed part'' and a  ``perturbation'' is not unique.
Prompted by the variational cluster approach (VCA), one
 can actually add ``auxiliary'' single particle terms to 
the unperturbed Hamiltonian 
and subtract them again within CPT.
This freedom can be exploited 
in order to ``optimize'' the results of the perturbative calculation.
As discussed in detail in Refs.~\onlinecite{da.ai.04,kn.ar.11},
in equilibrium this is an alternative way to motivate the
introduction of variational parameters in VCA. 
The  idea discussed here, thus, provides the natural extension of VCA to treat a
nonequilibrium 
steady state. There remains to define a criterion for the
``difference'' between initial and final state.
(Cluster) Dynamical Mean-Field Theory\cite{me.vo.89,ge.ko.96,ko.sa.01,ar.ko.11u} (DMFT) provides a natural solution,
requiring the cluster-projected Green's functions of the initial and
final state to coincide.
Of course, this self-consistency condition requires an infinite number
of variational parameters, as well as the solution of a (cluster)
impurity problem, which is computationally very expensive and whose
accuracy is limited, 
especially in real time. 
In equilibrium, the self-energy functional approach\cite{pott.03,pott.03.se} (SFA) provides one
possible generalization of DMFT if one wants to restrict to a finite
number of variational parameters. In this case, the requirement for the ``difference'' 
is provided by the Euler equation (see, \eg, Eq.~(7) in
Ref.~\onlinecite{pott.03}).

In the present paper, we  explore an
alternative criterion, represented by \eqref{vcond},
which, upon including an infinite number of bath sites,
becomes equivalent to  (cluster)-DMFT
 (see App.~\ref{appdmft}), similarly to SFA.~\cite{pott.03}
Without bath sites this corresponds to requiring that,
 for a given set of
variational parameters $\ve p$, their conjugate operators, \ie,
$d h/d \ve p$,
$h$ being the Hamiltonian, 
have the same expectation value in the unperturbed and in the
final target state. 
This criterion is numerically easier to implement than
the SFA, since in this case it is not necessary to search for a saddle point,
which is well known to be numerically expensive.~\cite{ne.se.08}
In addition, inclusion of bath sites  provides 
self consistency conditions for dynamic correlation functions as well.

The freedom discussed above
 can be additionally 
 exploited 
by including
 the hybridization between correlated regions and 
 the leads  as well as part of the leads themselves
 into the unperturbed Hamiltonian which is solved exactly by Lanczos
 exact diagonalization. In this way, CPT is then used to treat
 hopping terms 
further away
 from the correlated region.~\cite{central}
This partly accounts for the influence of the leads onto the
self-energy of the correlated region. 

Finally, let us mention that the method is probably most suited to deal with
models for which the correlated region is spatially
extended (see Fig.~\ref{laddergen}). In this case, this
region  must be partitioned into clusters which can be solved
exactly, while the intercluster terms are included into the
perturbative part.

\subsection{Variational cluster approach for nonequilibrium steady state}
\label{vcanss}
\begin{figure}
\begin{center}
 \includegraphics[width=0.45\textwidth]{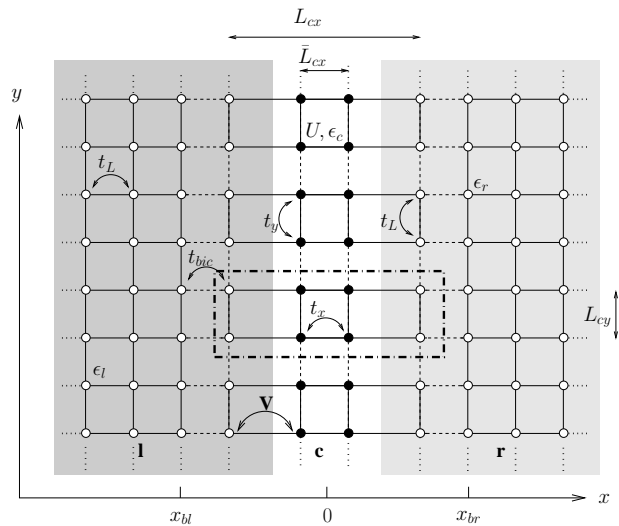}
\end{center}
\caption{\label{laddergen}
Generic scheme of the model studied here: 
full (empty) circles indicate correlated (uncorrelated) lattice sites.
Correlated sites define the correlated region (c), and
are characterized by an on-site Hubbard interaction $U$, an
on-site
energy $\epsilon_c$,  
and by hopping elements
$t_x$ and $t_y$ in the $x$ and $y$ direction, respectively.
The physical leads (l,r), indicated by the two shaded areas, consist of
half-infinite  planes described by uncorrelated tight-binding models
with hopping $t_L$,  on-site energies $\epsilon_l$ and $\epsilon_r$,
and chemical potentials $\mu_l$ and $\mu_r$, respectively.
The correlated region is connected to the leads via  
hoppings $V$. The width (number of sites in the $x$ direction) of the
 correlated region is $\bar L_{cx}$.
The height of the whole system in the $y$ direction is infinite.
In this
work, we study two 
cases, a strongly correlated chain ($\bar L_{cx}=1$) and a strongly correlated two-leg ladder ($\bar
L_{cx}=2$), both perpendicular to the
applied bias.
In the variational cluster calculation  the central 
region described by the unperturbed Hamiltonian $h$
can differ from the physical one. 
The latter coincides with the  correlated sites 
(white
area in the figure).~\cite{central}
On the other hand, 
the former consists of disconnected clusters
aligned along the $y$ direction, one
of them being represented by the dash-dotted rectangle in the figure. 
The corresponding equilibrium Green's function is determined  by
Lanczos exact
diagonalization. The size of these clusters is 
$L_c=L_{cx}\times L_{cy}$ ($4\times 2$ in the example).
The coordinates of the 
left and right boundary sites of the  central region
 are indicated by
$x_{bl}$  and $x_{br}$, respectively. 
Accordingly, dashed lines  represent hopping processes, which
are omitted in the  unperturbed (reference) Hamiltonian $h$
 and are re-included 
perturbatively within
 $\hat T$.
Full lines 
indicate hopping terms present in 
$h$, which are thus treated exactly (see text). 
}
\end{figure}
The physical model of interest consists of
a ``left'' and ``right'' noninteracting lead, as well as a
correlated region described by the Hamiltonians $\bar h_l$,
$\bar h_r$,
and $\bar h_c$, respectively, see \fig{laddergen}. 
$\bar h_c$ contains local (Hubbard-type)
interactions, as well as arbitrary single-particle terms.
For $\tau<\tau_0$, the three regions are  in equilibrium with three reservoirs at
different chemical potentials, $\mu_l$, $\mu_r$, and $\mu_c$
respectively. The correlated region is much smaller in size
than the leads, so that the latter act as  relaxation baths.
At $\tau=\tau_0$, the 
single particle (\ie, hopping) Hamiltonian
 terms $\hat{V}_{lc}$ and $\hat V_{rc}$
are switched on. These connect the left and right 
reservoir, respectively, with the correlated region.
The total time-dependent Hamiltonian is, thus, given by
\beq
H(\tau) = \bar h + \theta(\tau-\tau_0) \hat {\bar T}\;,
\eeq
where $\bar h = \bar h_c+\bar h_l+\bar h_r$, and
$\hat {\bar T} = \hat V_{lc}+\hat V_{rc}$.
We consider here the fermionic case, although many concepts can be
easily extended to bosons.
After a time $\Delta \tau$ long enough for relaxation to take place, the
system reaches a nonequilibrium steady-state, with a particle current flowing
from left to right for $\mu_l>\mu_r$ and from right to left for $\mu_l<\mu_r$. 

As discussed above, the total $\tau>\tau_0$ Hamiltonian  $H\equiv
H(\tau>\tau_0)$
is decomposed into an unperturbed part $h$ and a perturbation $\hat T$:
\beq
H = h + \hat T \;.
\eeq
In the simplest CPT approach for a ``small'' correlated region one can
take $h=\bar h$, and
$\hat T = \hat {\bar T}$. However, when the correlated region is
extended, as in Fig.~\ref{laddergen}, 
it has to be further decomposed into
smaller clusters that can be solved by exact diagonalization.~\cite{leads} 
In this
case, the intercluster hopping 
is subtracted from $h$ and must be included in $\hat T$.
In addition, one can include part of the leads into the clusters
(dashed lines in Fig.~\ref{laddergen}), so that 
$\hat V_{lc}+\hat V_{rc}$ are incorporated into $h$, while the leads
intercluster hoppings (e.g. $t_{bic}$ in the figure) are included~\cite{central} in
$\hat T$.
Finally, 
in the spirit of VCA, arbitrary intracluster terms $\Delta h$ can be added to the
unperturbed Hamiltonian and subtracted perturbatively within $\hat T$.
In other words, calling $h_{cl}$ the Hamiltonian 
describing the
physical cluster partition, and $\hat  T_{cl}$ the one
describing the intercluster hoppings (dashed lines in
Fig.~\ref{laddergen}),
we write $h=h_{cl}+\Delta h$, and
$\hat T = \hat T_{cl}-\Delta h$
so that the total Hamiltonian remains unchanged:
\beq
H = h_{cl} + \hat T_{cl}  = h + \hat T \;.
\eeq
The arbitrariness in the choice of $\Delta h$ can be exploited to
optimize the unperturbed state, as discussed in Ref.~\onlinecite{kn.ar.11}
for the equilibrium case.
Here, we will adopt a different optimization criterion, see discussion
below.
Being a single-particle term, $\hat T$ is described by its hopping
matrix $T$. This matrix has
a block structure according to the three regions discussed above
and shall be denoted by $T_{lc}$, $T_{rc}$ and $T_{cc}$, respectively.

Nonequilibrium properties, in general, and nonlinear transport in particular can
quite generally be determined in the frame of the Keldysh Green's
function approach.~\cite{kad.baym,schw.61,keld.65,ha.ja,ra.sm.86}
Here, we adopt the notation of Ref.~\onlinecite{ra.sm.86}, for which
the $2\times 2$ Keldysh Green's function matrix is expressed as
\beq
\label{G}
G(\vvr,\vvr'|\tau,\tau') = \left(\begin{array}{ll} G^R & G^K \\
                           0      & G^A 
\end{array}\right)
\;,
\eeq
where the retarded ($G^R$), advanced ($G^A$), and Keldysh ($G^K$)
Green's functions depend in general on two lattice sites ($\vvr,\vvr'$) and
two times ($\tau,\tau'$).
However,  both for $\tau<\tau_0$ as well as in steady state, time translation
invariance holds, so that Green's functions depend only on the time
difference $\tau-\tau'$, 
and we can Fourier transform to frequency space
$\omega$.

We use uppercase letters $G$ to denote Green's functions of the full
Hamiltonian $H$, and lowercase $g$ for the ones of the 
unperturbed Hamiltonian  $h$.
The advantage of using
the Keldysh Green's function matrix representation is that one can 
express Dyson's equation
in the same form as in equilibrium.~\cite{ha.ja,ra.sm.86}
In our case, we can express it in the form 
\beq
\label{dsigma}
G = g + g  \ (T+\Delta \Sigma) \ G  \;,
\eeq
where $g=\text{diag}\big(g_{ll},\,g_{cc},\,g_{rr}\big)$ is
  block diagonal,
and the products have to be considered as matrix multiplications.~\cite{freq}
In \eqref{dsigma}, $\Delta \Sigma = \Sigma -\Sigma_h$ is the
difference between the (unknown)  self-energy $\Sigma$
of the total Hamiltonian $H$, including the coupling to the leads, 
 and the self-energy 
$\Sigma_h$ associated with the unperturbed Hamiltonian $h$.

The CPT approximation\cite{se.pe.00} precisely amounts to neglecting $\Delta \Sigma$. As
pointed out in Ref.~\onlinecite{balzer_nonequilibrium_2011} this corresponds to neglecting
irreducible diagrams containing interactions and one or more $T$ terms.
It should, however, be
stressed that the self-energy of the
isolated clusters is exactly included in $g_{cc}$, which is obtained
by Lanczos exact diagonalization. 

In this approximation, \eqref{dsigma} can be used to obtain an
equation for the Green's function
$G_{cc}$
projected onto the central region, 
which is still a matrix in the lattice sites of the central
region and in Keldysh space~\cite{mnot}
(this is a straightforward generalization of, e.g., the treatment in Ref.~\onlinecite{ha.ja}): 
\begin{align}
\label{gcc}
G_{cc} &= g_{cc} + 
g_{cc} \big(\ T_{cc} \ G_{cc}+ \negthickspace\negthickspace
\displaystyle\sum_{\alpha}^{\in\{l,r\}}  T_{c\alpha} \ G_{\alpha c}\big) 
\intertext{and for the lead-central region Green's functions:}
\label{glc}
G_{\alpha c} &= g_{\alpha\alpha} \ T_{\alpha c} \ G_{cc}\;, \qquad
\text{with } \alpha\in\{l,r\}.
\intertext{
It is noteworthy that \eq{glc} is  exact and not based on the CPT approximation, as the leads contain non-interacting particles. Insertion of \eqref{glc} into \eqref{gcc} yields}
G_{cc} &= g_{cc} + g_{cc} \big(T_{cc} + \tilde \Sigma_{cc} \big)\ G_{cc}
\label{eq:gcc}
\intertext{with the lead-induced self-energy renormalization}
\label{sigmaeff}\tilde \Sigma_{cc} &=  
\displaystyle\sum_{\alpha}^{\in\{l,r\}} 
T_{c\alpha} \ g_{\alpha\alpha}  \ T_{\alpha c}\;.
\end{align}
Here $g_{\alpha\alpha}$ stands for the Green's function of the isolated lead $\alpha$.
One finally obtains a Dyson form
for the steady state Green's function of the coupled system at the central region
\beq
\label{gcccpt}
G_{cc}^{-1} = g_{cc}^{-1}  - T_{cc} - \tilde \Sigma_{cc} \;.
\eeq
Different from the usual Dyson equation,
$g_{cc}$ is  the Green's function for the isolated clusters, which
contains all many-body effects 
inside the cluster.

For the evaluation of the current from, say, the left lead to the
central region~\cite{central} one needs 
the $G_{lc}$ Green's function, which is readily obtained by combining
\eqref{glc} with \eqref{gcccpt}.
This leads to the generalized Kadanoff-Baym equation (see \eg Refs.~\onlinecite{ha.ja,me.wi.92}),
along with the fact that the central region is finite in $x$ direction and the leads are infinite, 
one can rewrite 
the current into a B\"uttiker-Landauer type of formula
\begin{align}
j = \int \frac{d\varepsilon}{2\pi}&
\big[
f_{F}(\varepsilon-\mu_{r})-f_{F}(\varepsilon-\mu_{l})
\big]\nonumber\\
&\times\Tr\big[
\Gr{cc}{}(\varepsilon) \Gamma^{}_{l}(\varepsilon) \Ga{cc}{}(\varepsilon) \Gamma^{}_{r}(\varepsilon)
\big]\label{eq:meir:wingreen}\;.
\end{align}
where $G^{R/A}_{cc}$ is the retarded/advanced part of the Green's
function $G_{cc}$,
 and the trace, as  well as matrix products run over site indices in $c$.  
 $\Gamma_{\alpha}$ describes the inelastic broadening owing to  the 
 coupling to lead $\alpha$, which in CPT is given by~\cite{real}
\begin{align*}
\Gamma_{\alpha} &= 2 \im \big\{
T_{c \alpha} g^{A}_{\alpha\alpha} T_{\alpha c}
\big\}\;,
\end{align*}
which represents the contribution of lead $\alpha$ to the imaginary part of $\tilde \Sigma_{cc}^{A}$.
Interestingly, the expression for the current in CPT  has the same
structure as  the Meir-Wingreen formula \cite{me.wi.92} for
non-interacting particles, 
which is the basis for 
nonequilibrium ab-initio-calculations.\cite{fu.br.08}
Here, however, the Green's function contains the many-body interactions of the correlated region.
 An advantage of this expression is that it 
yields an explicit connection to the Green function $G^{R/A}_{cc}$ of the scattering region and 
the influence of the itinerant electrons in the leads.
A similar expression can be derived for the one-particle density matrix 
between two sites with the same $y$ coordinate,
which is required for the self-consistency condition discussed below.

As it is well known,  
 all  retarded and advanced Green's
functions  are evaluated without chemical potentials. The latter enter
through the Keldysh Green's function or rather via the Fermi
functions. 
While the chemical potential of the central region  is wiped out in
the steady state due to its small size in comparison to the size of the leads,  
the chemical
potentials of the leads explicitly enter the expressions for the
current and the density matrix, see \eq{eq:meir:wingreen}.  
In the case investigated here, 
the central region is translation invariant in $y$ direction and is split 
into
identical clusters.
In the end, as far as the main numerical task is concerned, one has to solve
many-body problems for clusters of size $L=L_{cx}\times L_{cy}$,
invert matrices of the same size, and sum over wave vectors 
$q_y$ belonging to the Brillouin zone associated with the cluster supercell.

\subsection{
Self-consistency condition}
\label{sc}
Equation \eqref{gcccpt} is the expression for the Green's function of
the central region 
within the CPT approximation. 
As discussed above, one would like to
optimize 
the initial state 
in some appropriate way
by suitably adjusting the parameters 
$\Delta h$
of the
unperturbed Hamiltonian $h$.
 The inclusion of
additional terms $\Delta h$ adds flexibility to 
the
self-energy $\Sigma_h$ which is included within this approximation.
 Obviously, it makes no difference in the case of
non-interacting particles as the selfenergy vanishes exactly,
independently of $\Delta h$.
 This
freedom can be exploited in order  to improve 
the approximation systematically.
A similar discussion on this issue has been given in
Refs.~\onlinecite{da.ai.04,kn.ar.11}), 
and is at the basis of the VCA idea~\cite{pott.03}. 

As discussed above, we need a variational condition associated with
a ``minimization'' of the difference between unperturbed and perturbed
state. In (cluster)-DMFT one requires the cluster projected Green's
function  to be equal to the unperturbed one 
\beq
\label{gdmft}
g_{cc} = \pcluster(G_{cc}) \;,
\eeq
where $\pcluster$ projects the Green's function onto the cluster,
\ie, it sets all its intercluster matrix elements to
zero.~\cite{coincide}
Since here we have a finite number of variational parameters $\ve p$ 
that can
be adjusted, 
we cannot satisfy \eqref{gdmft}. We, thus, propose a ``weaker'' condition, namely 
 that the 
expectation values of 
operators coupled to the variational parameters 
contained in $\Delta h$  
(\ie, $d \Delta h/d \ve p$) 
be equal 
in the unperturbed and in the perturbed state.
More specifically, we impose the condition
\beq
\label{vcond}
\int \frac{d\omega}{2\pi}\tr  
\hat{\tau}_1 \frac{\partial \left(g_{0cc}\right)^{-1}}{\partial \ve p} 
\left(g_{cc}-G_{cc}\right)
= 0 \;,
\eeq
where $\hat{\tau}_1$ is a Pauli matrix in Keldysh space,~\cite{tau1}
and $g_{0cc}$ is the Green's function associated with the noninteracting part of $h$. 
It is interesting to note
(see appendix ~\ref{appdmft})
 that 
by including into $\Delta h$ a coupling to an infinite
number of bath sites,
the present method, with the self-consistence condition 
\eqref{vcond} whereby
$\ve p$ are the bath parameters
(hopping and on-site energies),
 becomes equivalent to nonequilibrium cluster DMFT.
Generalization of the SFA condition to nonequilibrium should be, in
principle, obtained by replacing $g_{0cc}$ with $\Sigma_h$ in \eqref{vcond}.

A second systematic improvement 
of this nonequilibrium VCA approach
consists in increasing the cluster
size $L_c$. This can be done in two ways: (i) by extending the
boundaries of the central region in $y$ direction and thus treating
more correlated sites exactly~\cite{central} and (ii) by extending the boundaries in
$x$ direction  
to  
include
an increasing number of uncorrelated lattice sites, \ie, taking
$L_{cx}>\bar L_{cx}$, \textit{cf.} Fig.~\ref{laddergen}.
This amounts to taking into account to some
degree the $V$-induced renormalization of the self-energy.

\section{Model}
\label{model}
\begin{figure*}
\begin{center}
 \includegraphics[width=\textwidth]{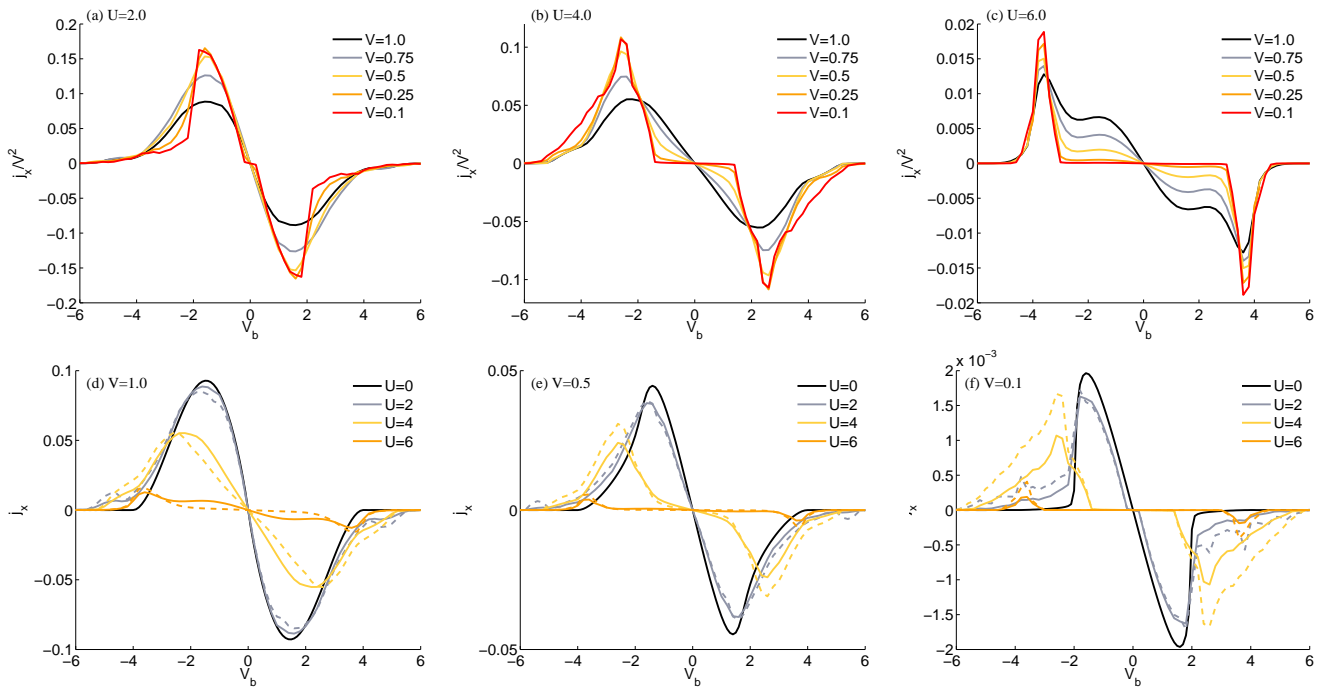}
\end{center}
\caption{\label{jladder}
(Color online)
Steady-state current density $j_x$ versus bias voltage $V_b$
for a correlated two-leg ladder ($\bar L_{cx}=2$). First row shows
 $j_x$  normalized by $V^2$ as function
of $V_{b}$ evaluated for different values of $V$ and of the
interaction \fc{a} $U=2.0$, \fc{b} $U=4.0$, and \fc{c} $U=6.0$.  Second row shows the
$U$ dependence of the current for different values of the
 hopping 
$V=V_{lc}=V_{rc}$
from
the leads to the correlated region
  \fc{d} $V=1.0$, \fc{e} $V=0.5$, and \fc{f} $V=0.1$. Solid (dashed)
  lines represent results for the current 
between the left lead and the central region (between two in $x$ direction adjacent sites inside the central region), \ie,
evaluated with $G_{lc}^K$ ($G_{cc}^K$), see text for details.
Results are obtained 
 by using  a reference Hamiltonian $h$ consisting
of disconnected
 clusters of size $L_c=L_{cx}\times
L_{cy}=2\times6$.
}
\end{figure*}

Next, we present an application of the nonequilibrium VCA method described in Sec.~\ref{method}.
Specifically, we study nonlinear transport properties
across an extended correlated region (denoted as $c$ in Fig.\ref{laddergen}), which
we take to be a Hubbard 
chain ($\bar L_{cx}=1$)
or a Hubbard ladder ($\bar L_{cx}=2$) with nearest-neighbor hoppings
$t_x$ and $t_y$, on-site interaction $U$, on-site energy
$\epsilon_c$, and chemical potential $\mu_c$
\[
\bar h_c=\sum_{\langle i,j \rangle,\,\sigma} t_{ij} c_{i\sigma}^\dag c_{j\sigma}^\nag + U \sum_{i} \hat{n}_{i\uparrow} \hat{n}_{i\downarrow} + (\epsilon_c - \mu_c)\sum_{i,\sigma}\hat{n}_{i\sigma} \;,
\]
in usual notation, and
where $t_{ij}=t_x$ ($t_{ij}=t_y$) for $i$ and $j$ being nearest neighbors in $x$ direction ($y$ direction).
The leads (shaded regions in Fig.~\ref{laddergen}) 
are described by 
two-dimensional semi-infinite tight-binding models
 with nearest-neighbor hopping $t_L$, on-site
energies $\epsilon_l$ and $\epsilon_r$, and chemical
potentials $\mu_l$ and $\mu_r$ for the left and right lead, respectively. 
We apply a bias voltage $V_b$ to the leads by setting $\mu_{r} =  \epsilon_{r}=V_b/2$ and restrict to the particle-hole symmetric case where
$\epsilon_c=-U/2$, $\mu_c=0$, $\epsilon_r=-\epsilon_l$, and
$\mu_l=-\mu_r$. 
For simplicity, we neglect the long-range part of the Coulomb
interaction. Under some conditions, this can be absorbed
within the single-particle parameters of the Hamiltonian, in a
mean-field sense.~\cite{ha.ja} 

As discussed above, the unperturbed Hamiltonian $h$ does not necessarily
coincide with the physical partition into leads and 
correlated region. $h$ is obtained by tiling the total system into small clusters as
illustrated in Fig.~\ref{laddergen}, as well as by adding an
intracluster variational term $\Delta h$.

In the present work $\Delta h$ describes
a correction $\Delta t_x$ 
to the intra-ladder hopping.
Further options could include, for instance, a site-dependent change in
the on-site energy $\Delta \epsilon_c(x)$. Particle-hole symmetry can be preserved by
constraining this change to be antisymmetric: 
$\Delta \epsilon_c(x)= -\Delta \epsilon_c(-x)$.
In this paper, whose goal is to carry out a first test of   the method, 
we restrict, for simplicity, to a single variational parameter.
The choice of $\Delta t_x$ as a variational parameter is motivated by
the fact that this term is important for the current flowing in $x$ direction.
According to the prescription discussed above, we require the
expectation value of the 
one-particle density matrix for nearest-neighbor indices in $x$
direction to be the same
for the unperturbed $h$ and for 
the full $H$, i.e.  evaluated with $g_{cc}$ and with $G_{cc}$.

One comment about the chemical potential. In principle, when 
including
some of the sites of the leads in $h$, \ie, when
$L_{cx}>\bar L_{cx}$, then these additional sites have a chemical
potential $\mu_c$ which differs from 
the one they would have if
$L_{cx}=\bar L_{cx}$ (\ie, $\mu_l$ or $\mu_r$). However, the chemical potential, of these sites  does not affect the
steady state, as their 
volume-to-surface ratio is finite.
Of course, their on-site energies ($\epsilon_r$ and $\epsilon_l$) 
are important.

Due to translation invariance by a cluster length
$L_{cy}$ in the $y$ direction, 
it is convenient, as in usual VCA, to carry out a 
Fourier transformation in $y$ direction, with associated momenta $q_y$.
The Green's functions $g_{cc}$ and $G_{cc}$, as well as $T$
become now functions of two momenta $q_y+Q_y$ and $q_y+Q_y'$, where
$Q_y$ and $Q_y'$ are reciprocal superlattice vectors of which there
are only $L_{cy}$ inequivalent ones. 
In order to evaluate the nonequilibrium steady state, one 
only needs the equilibrium Green's function
$g(x_{b\alpha}|q_y|z)$
 of the isolated leads at the contact edge
to the central region, with $x$ coordinate equal to $x_{b\alpha}$
($\alpha \in \{l,\,r\}$), and Fourier transformed in the $y$ directions, where
$q_y$ is the corresponding momentum and $z$ the complex frequency.
For a semiinfinite nearest-neighbor tight-binding plane with hopping
$t_L$, and on-site energy $\epsilon_{\alpha}$, this
can be expressed as
\beq
\label{gxb}
g(x_{b\alpha}|q_y|z) = g_{c,loc}(z-2 t_L  \cos q_y-\epsilon_\alpha)\;,
\eeq 
where
 $g_{c,loc}(z)$ is the local Green's function of
a tight binding chain with open boundary conditions and with zero
on-site energy. The latter can be  determined 
analytically along the lines discussed in
Ref.~\onlinecite{economou_greens_2006}.

The model studied here,  is motivated by the interest in transport
across semiconductor heterostructures
(see, e.g. \onlinecite{pe.gu.08,ertler_proposal_2010,ertler_self-consistent,ch.le.11}).
However, it is well known that in this case charging effects are
important, also near the boundaries between the leads and the central
region.
Here, scattering effects produce charge density waves, which, when
taking into account the long-range part of the Coulomb interaction,
even in mean-field, produce a modification of the single-particle
potential.
In order to treat realistic structures, these effects should be
included at the Hartree-Fock level at least. All these generalizations
can be 
straightforwardly treated with the presented variational cluster
method, however, in this work we focus on a first proof of concept study and application containing
the essential ingredients for the investigation of the nonequilibrium steady state of strongly correlated many-body systems.
\begin{figure*}
\begin{center}
 \includegraphics[width=\textwidth]{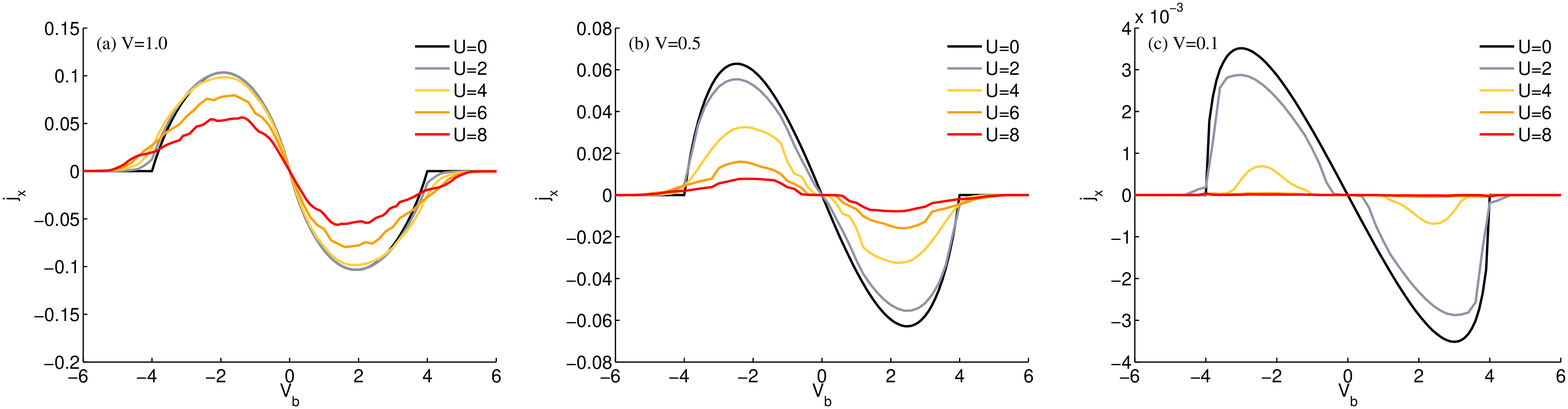}
\end{center}
\caption{\label{jchain}(Color online)
Steady-state current density $j_x$ as in Fig.~\ref{jladder} but
for the correlated chain ($\bar L_{cx}=1$). The current density is evaluated for different values of
the lead to correlated region hopping \fc{a} $V=1.0$, \fc{b} $V=0.5$, and \fc{c} $V=0.1$, and of the interaction $U$, see legend.
Results are obtained for reference clusters 
 of size $L_c=L_{cx}\times L_{cy}=3\times 4$. 
}
\end{figure*}

\section{Results}
\label{results}

We have evaluated the steady-state current density $j_x$ of the models discussed in
Sec.~\ref{model} as a function of the bias 
$V_b\equiv \epsilon_{r}-\epsilon_{l}$
between the leads at zero temperature.
Simultaneously the chemical potential is adjusted
to the on-site energy $\mu_{\alpha} = \epsilon_{\alpha}$, which
corresponds to a rigid shift of the density of states in both leads in
opposite directions.

In \fig{jladder} we display results for the
two-leg ladder ($\bar L_{cx}=2$),
for different values of the interaction strength
$U=\lbrace 0,\,2,\,4,\,6\rbrace$ and lead-to-system hopping
$V=\lbrace1.0,\,0.75,\,0.5,\,0.25,\,0.1\rbrace$. 
We use $\hbar=1$ and $t_L=1$ which sets the unity of
energy. 
Moreover, we take the lattice 
constant $a=1$. The hopping is uniform in the whole system, meaning
that $t_x,\,t_y$ in the correlated region and $t_L$ of the leads
are equal. The on-site energy of the
correlated region is $\epsilon_c=-U/2$ corresponding to half-filling,
whereas the on-site energy of the left (right) lead is
equal to its chemical potential
 $\mu_l$ ($\mu_r$). The 
unperturbed hamiltonian $h$ describes the central region  decomposed into clusters of
 size $L_c=2\times6$. 
The corresponding Green's function $g_{cc}$ is determined exactly by
 Lanczos  diagonalization.
All results are determined self-consistently using $\Delta t_x$ as
variational parameter, see Sec.~\ref{model}.

Using the Meir-Wingreen expression, \eq{eq:meir:wingreen}, the general
trend of the results for the steady-state current $j_x$ can be
discussed conveniently. At zero temperature there are only
contributions to the current for $\min(\mu_l,\mu_r) < \omega <
\max(\mu_l,\mu_r)$ due to the difference of the Fermi distribution
functions. In particular this leads as expected to zero current for
zero bias voltage $V_b$. 
With increasing bias voltage $V_b$ the modulus of $j_x$ initially increases. 
For large values of $V_{b}$ it decreases again, as
the overlap of the local density of states of the 
two leads enters
the expression,
 which is zero if $V_{b}$ is greater than the band
width of the leads. Hence the local density of states of the leads
along with the Fermi function act as a filter that averages the
electronic excitations of the central region within a certain energy
window.  \begin{SCfigure*}
\centering
 \includegraphics[width=0.7\textwidth]{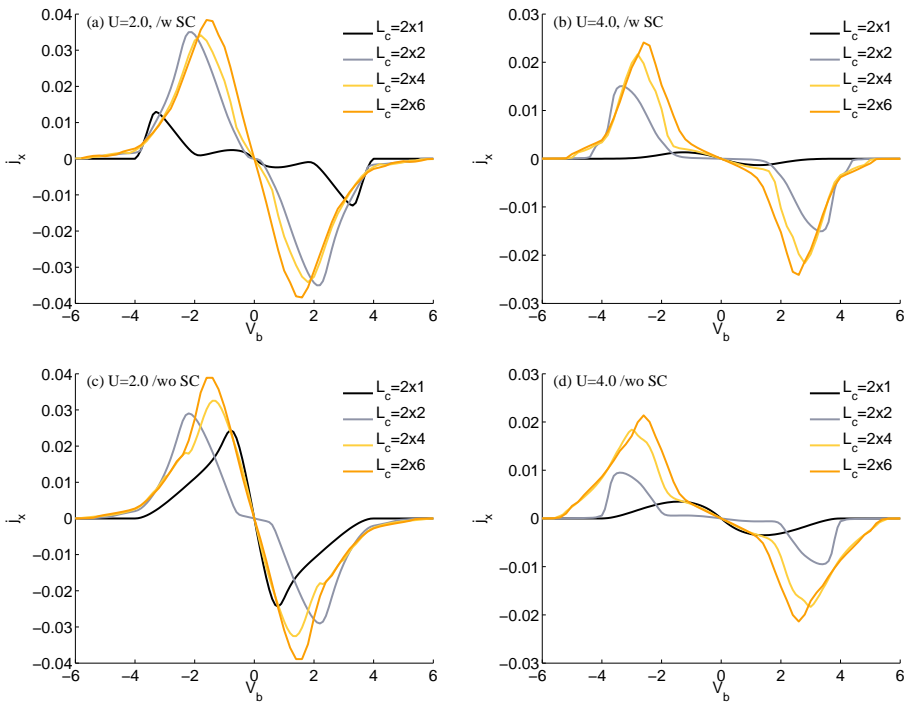}
\caption{\label{convLadder} (Color online)
Convergence of the steady-state current density $j_x$ with 
reference cluster 
size $L_c=L_{cx}\times L_{cy}$ for the correlated two-leg
ladder with $V=0.5$. Results in \fc{a}, \fc{b} are obtained by a
  variational adjustment of the intra-cluster hopping $t_{x}$ as discussed in the text,  
while those of \fc{c}, \fc{d} are obtained without modification of
$t_{x}$. The values for the Hubbard interaction are $U=2$ in [\fc{a}, \fc{c}] and $U=4$ in  [\fc{b}, \fc{d}].
}
\end{SCfigure*}

In the system we are studying, the leads are modeled by semi-infinite tight binding planes.
Alternatively, instead of using \eqref{gxb} one could simply put a model
Green's function ``by hand,''
as for example one which describes a
 Lorentzian shaped density of states.
 Such an unbound density of states generally leads to 
a finite value of the current for arbitrary bias.

The leads have a further effect on the result as they provide an
inelastic broadening of the energy spectrum of the central region
entering $\Sigma^{\text{eff}}$, see \eq{gcccpt}, which smears out
details of the excitation spectrum.
As far as the lead-correlated region coupling $V$ is concerned,
there are two competing effects: on the one hand,  
 the 
current increases with increasing $V$ due to the 
stronger coupling between the correlated region and the leads.
On the other hand, details of the electronic excitations are smeared
out with increasing $V$ leading to a reduced resolution.
Therefore,
in order to detect the effects of strong correlations,
particularly the gap, a small value for $V$ is required.

The details of the $V$ dependence 
of $j_x$ for small $V$
can be deduced from
\eqref{eq:meir:wingreen}.
Here, the expression for the current has a prefactor
proportional to $V^{4}$ (at least in the $L_{cx}=\bar L_{cx}$ case),
due to the two $\Gamma$ terms.
On the other hand, for a gapless system,
 there is a $V^{2}$ term in the denominator of 
$|G_{cc}^R|^{2}$.
For a gapped system, this is cut off by the energy gap $E_g$, so that 
in this case $j_x\sim V^{4}/E_{g}^{2}$, while 
$j_x\sim V^2$ for a gapless spectrum.
These aspects are clearly observable in \fig{jladder} \fc{a}--\fc{c}, 
which shows the scaled  
current density $j_x/V^2$ for fixed interaction
strength $U$ but 
varying $V$. 
The envelope
has a rotated S-like structrue due to the 
combined effects of the lead density of states and of the Fermi  
functions. 

% MKCH
% The overall size of the 
% structures deceases with increasing Hubbard interaction 
% $U$ and the gap becomes more pronounced with decreasing $V$.

Next we will analyze a bit more in detail \nomkch{the effects of
the Hubbard interaction.} % the results for the % MKCH
%current across the Hubbard ladder in \fig{jladder}.
Increasing the interaction strength $U$ in the correlated region leads to
a suppression of the current and the opening of a gap, which is best
oberserved in \nomkch{\fc{f}}. For $U=4$ the maximum of the current density is roughly reduced by a
factor of two as compared to the noninteracting case, whereas for
$U=6$ the current is almost one order of magnitude smaller as compared
to the noninteracting system, see \fig{jladder} \fc{d}--\fc{f}.  

Finally, we want to address the the difference between 
the solid lines and dashed lines in the the panels \nomkch{\fc{d}--\fc{f}}
of \fig{jladder}, which represent the current density 
evaluated on a bond connecting the leads to the central region, or 
on a bond within 
the two-leg ladder.
Due to the stationary condition, the two results should coincide.
However, our calculations shows  
a slight discrepancy between 
them, which is  due to the fact
that the method is not completely conserving and, thus, the continuity
equation is not completely fulfilled. However, from our results we
see that the deviation from the continuity equation is quite small. 
We expect this discrepancy to be reduced upon improving the
optimization with the introduction of additional variational parameters.

In Fig.~\ref{jchain} we show the steady-state current density $j_x$
across the correlated chain ($\bar L_{cx}=1$) as a function of the bias
voltage. The parameters are the same as in the case of the two-leg
ladder, however, the central region is decomposed into clusters of
size $L_c=3\times4$, where also sites of the leads are taken
into account to improve the results. 
The half-filled Hubbard chain is gapped as well. As for the two-leg
ladder, the gap behavior can be better seen in the current-voltage characteristics
 for smaller values of $V$, in our case for $V=0.1$.
 In contrast, for strong coupling $V=1.0$, \fc{a}, 
no gap behavior can be seen in the current
 due to the strong hybridization with
the leads.

For strong values of the coupling $V$ between leads and correlated region ($V=1.0$),  
\fc{a}, the
current is significant 
for all values of the interaction $U$.
 However, with decreasing $V$, \fc{b}--\fc{c}, the current is
 strongly suppressed for large interaction $U$. 
 Importantly, for the
 correlated chain the continuity equation is always strictly
 fulfilled. In other words, there is no difference between $j_x$ evaluated
on a intercluster bond between the leads and the cluster, or on an
intracluster bond. 
This is  due to the absence of vertex corrections at the uncorrelated sites.

Next, we study the convergence 
of our results with the size of the  cluster,
as well as the effect of the self-consistency condition
for the two-leg ladder and $V=0.5$. Results are depicted  in Fig.~\ref{convLadder} 
for two different values of the Hubbard interaction, namely  $U=2$
[\fc{a}, \fc{c}] and $U=4$ [\fc{b}, \fc{d}]. 
We do not plot results of the convergence analysis 
for $U=6$, since for this large $U$ the current is already rather small,
as can be  seen in \fig{jladder} \fc{d}--\fc{f}.
Results in \fc{a} and \fc{b}, first row, are obtained 
by adjusting $\Delta t_x$ self-consistently, as described in Sec.~\ref{model},
whereas \fc{c} and \fc{d}, second row, shows
results without self-consistency, \ie, with $\Delta t_x=0$.
Results show that 
the self-consistency procedure 
improves the results, 
as the convergence for $j_x$ 
is faster with increasing
cluster size as compared to the case without self-consistency. Generally, we observe pronounced finite size effects for very small clusters up to $2\times 4$, and convergence seems to be reached for the $2\times 6$ cluster.

\begin{figure}
\begin{center}
 \includegraphics[width=0.4\textwidth]{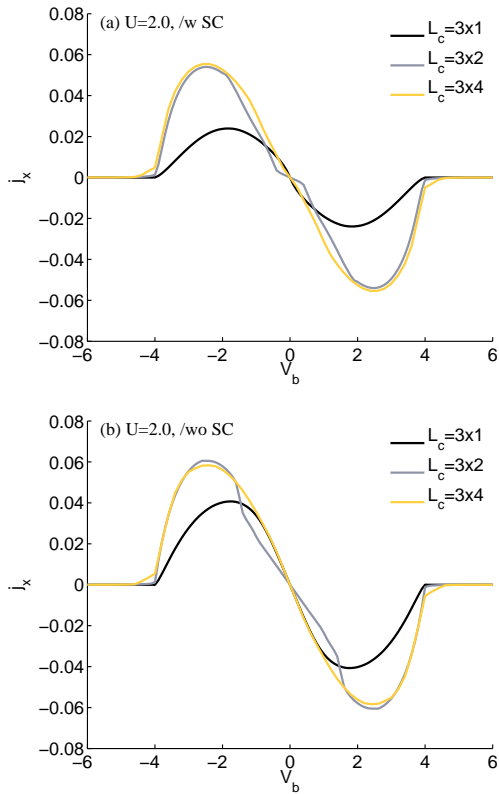}
\end{center}
\caption{\label{convChain} (Color online)
Convergence of the steady-state current density $j_x$ with
reference cluster size $L_c=L_{cx}\times L_{cy}$ for the correlated chain. Results in \fc{a} fulfill the self-consistency condition \eqref{vcond}, whereas results in \fc{b} do not. The parameters are $U=2$ and $V=0.5$.
}
\end{figure}

We now repeat the same analysis for the correlated chain. The corresponding current densities for the parameters $U=2$ and $V=0.5$ are shown in
\fig{convChain} for different cluster sizes.  
Results shown in \fc{a} are with self-consistency procedure \eqref{vcond}, whereas the results shown in \fc{b} are without. 
In the present case, where we consider transport across a 
strongly correlated chain, convergence is achieved very quickly with
increasing cluster size.
Therefore, there is no sensible difference between results obtained
with or without self-consistency, apart for the pathological case
$L_c=3\times1$ (see below).

Results obtained for the two-leg ladder and for the chain show that cluster geometries with $L_{cy}=1$ provide results far from convergence, even with self-consistency.  
For the chain this is probably due to the degeneracy 
of the cluster ground state. For the ladder, it seems that using as
starting point the $2\times1$ dimer exaggerates the gap. 
But besides these data obtained from admittedly very small clusters, results converge quickly as a function of
cluster sizes, especially when the hopping in $x$ direction is used as a variational parameter.

\section{Conclusions}
\label{conclusions}

In this paper we have presented a novel approach to
treat  strongly correlated systems in the nonequilibrium
steady state. The idea is based on the 
variational cluster approach extended to the Keldysh formalism.
For the present approach
the expression for
the current resembles the corresponding Meir-Wingreen formulas.
As in the original Meir-Wingreen approach, which is also the basis for
nonequilibrium density-functional based 
calculations, we directly address the
steady state behavior  of a device coupled to infinite leads. The
latter is necessary for 
the system to reach a well-defined steady state. 

The present nonequilibrium extension is in a similar spirit 
to the equilibrium self-energy functional approach, in which one ``adds'' single-particle terms to the 
cluster
Hamiltonian which is then solved exactly, and ``subtract'' them
perturbatively.~\cite{da.ai.04,kn.ar.11} The  values of the
parameters are determined by an appropiate requirement which in the end
amounts to optimizing the unperturbed state with respect to the
perturbed one. 

There is a certain freedom in choosing the most appropriate
self-consistency criterion. Here we have required the operators associated with
the variational parameters to have the same expectation values in the
unperturbed and in perturbed state. 
Certainly, an interesting alternative would be to generalize the 
variational criterion provided by the 
self-energy functional approach~\cite{pott.03} to the nonequilibrium case.
This will be obtained by a suitable generalization of 
the Euler equation (Eq. (7) of Ref.~\onlinecite{pott.03})
 to the Keldysh contour, \ie, by replacing 
$g_{0cc}$ with the self-energy $\Sigma_h$ in \eqref{vcond}
Work along these lines is in progress.

The advantage of the present variational condition \eqref{vcond} is
that it is computationally less demanding, as one just needs to
evaluate cluster single-particle Green's functions.
Which one of the two conditions gives more accurate results cannot be
stated a priori and should be explicitly checked.

In any case, both methods, the self-energy functional approach and the present
one,  become equivalent to
(cluster) dynamical mean-field theory 
in the case in which an infinite number of variational
parameters is suitably taken
(see Appendix~\ref{appdmft}).

In general, we  expect results to improve when more variational
parameters are taken into account. 
In particular,
when evaluating the current across the central region, it would be
useful if a current was already flowing in the cluster. This can be
achieved by adding a complex variational hopping between the end points
of the cluster, and of course remove it perturbatively. 
The corresponding variational condition would contain the 
interesting requirement that the current flow in this modified cluster be the same as in
steady state.

The model studied here,  is motivated by the interest in transport
across semiconductor heterostructures
(see, e.g. \onlinecite{pe.gu.08,ertler_proposal_2010,ertler_self-consistent,ch.le.11}).
However, it is well known that in this case charging effects are
important, also near the boundaries between the leads and the correlated
region.
Here, scattering effects produce charge density waves, which, when
taking into account the long-range part of the Coulomb interaction,
even in mean-field, produce a modification of the single-particle
potential.
In order to treat realistic structures, these effects should be
included at the Hartree-Fock level at least. All these generalizations
can be 
straightforwardly treated with the presented variational cluster
method, however, in this work we focus on a first proof of concept study and application containing
the essential ingredients for the investigation of the nonequilibrium steady state of strongly correlated many-body systems.

\section*{Acknowledgments}

We acknowledge fruitful discussions with 
S.~Diehl.
We made use of the ALPS library.\cite{albuquerque_alps_2007}
Financial support from 
the Austrian Science Fund (FWF)
under Projects No. P18551-N16,
and P21289-N16 are gratefully acknowledged.

\appendix

\section{Connection to (cluster) Dynamical Mean-Field Theory}

\label{appdmft}
Here, we show that the self-consistent condition \eqref{vcond} provides a bridge to
(cluster) DMFT, when an increasing number of noninteracting bath sites
\noeach{%CCEA
with appropriate parameters and occupations %CEA
}%CCEA
is  included in the  cluster Hamiltonian. Notice that these are ``auxiliary''
baths and are not related to the leads.
Concretely, 
\noeach{%CCEA
this is achieved by introducing into the
 variational Hamiltonian
\nomkch{$\Delta h$}
a coupling  of the central region  with a set of uncorrelated bath
sites with appropriate energies, hybridizations $v^{\pm}_n$, and
occupations (see below).
The hybridizations $v^{\pm}_n$ and the energies 
are therefore  ``included'' in $h$, but 
``subtracted perturbatively,'' from 
the target Hamiltonian $H$. Their parameters are determined
variationally via \eqref{vcond}.
}%CCEA

Now, since  $g_{0cc}$ is cluster-local, a solution to \eqref{vcond} 
is obviously given by \eqref{gdmft}. However, this solution can
generally not be obtained with a finite number  
of parameters. As in usual equilibrium (cluster) DMFT,~\cite{ge.ko.96}
\eqref{gdmft} can thus be 
solved via an iterative procedure defined by
\beqn
g_{0cc,new}^{-1} &&= \left(\pcluster(G_{cc})\right)^{-1} + \Sigma_h 
\nonumber \\ 
\Sigma_h &&= g_{0cc,old}^{-1} - g_{cc}^{-1}  \;.
\eeqn
It is, thus, sufficient to show that an arbitrary 
$g_{0cc,new}$ can be obtained by coupling the cluster to a 
noninteracting bath with suitably chosen bath parameters.
For the retarded and advanced Green's functions, the procedure is the
same as in equilibrium. The Keldysh part is slightly more complicated.
In order to show that an aribrary $g_{0cc,new}$ can be realized, one introduces
 the hybridization function
\beq
\label{deltaom}
\Delta(\omega)= \left(\begin{array}{cc}
\Delta^{R}(\omega) &  \Delta^{K}(\omega) \\
                           0      & \Delta^{A}(\omega) 
         \end{array}\right) \;,
\eeq
where the $\Delta^{R}$, $\Delta^{A}$, and $\Delta^{K}$
are matrices in the
 cluster sites.
Similarly to equilibrium DMFT
$g_{0cc,new}$ is expressed as
\beq
g_{0cc,new}^{-1} =  g_{0cc,0}^{-1} - \Delta(\omega) \;.
\eeq 
Here, $g_{0cc,0}^{-1}$ is the ``bare'' noninteracting cluster Green's
function, \ie, the one
with neither baths nor variational parameters.

\noeach{%CCEA
An arbitrary (steady-state) $\Delta(\omega)$ can be produced by 
coupling the central region to
an appropriate bath  in the
following way.
The retarded (and advanced) part are obtained as in equilibrium DMFT~\cite{ge.ko.96} 
by coupling to a bath with spectral function~\cite{imre} $A_{bath}(\omega)$ 
given by
\beq
\label{abath}
  \ A_{bath}(\omega) = 
- \frac{1}{\pi}\im \Delta^{R}(\omega) \;,
%-2 i \pi \ A_{bath}(\omega) = 
%\left(\Delta^{R}(\omega)-\Delta^{A}(\omega)\right)
\eeq
($\re\Delta^R$ is fixed by Kramers-Kronig relations). 
%%\mkcomm{Kronig?, laut wikipedia Kronig, ich habe aber auch schon Kr\"onig geh\"ort}
On the other hand, the Keldysh part is generated by splitting the bath
defined by \eqref{abath} into two baths, a %MKCH full
full ($\mu=\infty$) and an empty 
($\mu=-\infty$) one, respectively. Their spectral functions are denoted by
$A_{bath}^+(\omega)$ and $A_{bath}^-(\omega)$, respectively, and
should obviously fullfill the condition
\beq
\label{abathsum}
 A_{bath}(\omega) = A_{bath}^++A_{bath}^- \;.
\eeq
Since the Fermi functions of the two baths are $1$ and $0$, respectively,
the Keldysh part $\Delta^{K}(\omega)$ is given by
($\Delta^K$ is anti-hermitian)
%%\mkcomm{Neuer Buchstabe f??r $A^K_{bath}$, zB $B_{bath}$? Verstehe ich das richtig?}
\beq
\label{ak}
\Delta^{K}(\omega) = -2 i \pi \ B_{bath}(\omega) \equiv   
-2 i \pi \ \left(A_{bath}^-(\omega)-A_{bath}^+(\omega)\right)   \;.
\eeq
This fixes the two spectral functions to be\nomkch{
\beq
A^\mp_{bath}(\omega) = \frac{A_{bath}(\omega)\pm B_{bath}(\omega)}{2}\;.
\eeq}
As usual, the two baths spectral functions 
$A_{bath,\pm}(\omega)$
are realized
by coupling the central region with a set of noninteracting sites with
energies $\epsilon^{\pm}_n$
and hybridizations~\cite{imre} $v^{\pm}_n$, fixed by\nomkch{
%The spectral functions are then
\beq
A_{bath}^\pm(\omega) = \sum_n v^{\pm}_n v^{\pm\dag}_n \delta(\omega - \epsilon^{\pm}_n)\;.
\eeq
}%CCEA
}

\bibliographystyle{apsrev}
\bibliography{references_database,library,libraryea}

\begin{thebibliography}{72}
\expandafter\ifx\csname natexlab\endcsname\relax\def\natexlab#1{#1}\fi
\expandafter\ifx\csname bibnamefont\endcsname\relax
  \def\bibnamefont#1{#1}\fi
\expandafter\ifx\csname bibfnamefont\endcsname\relax
  \def\bibfnamefont#1{#1}\fi
\expandafter\ifx\csname citenamefont\endcsname\relax
  \def\citenamefont#1{#1}\fi
\expandafter\ifx\csname url\endcsname\relax
  \def\url#1{\texttt{#1}}\fi
\expandafter\ifx\csname urlprefix\endcsname\relax\def\urlprefix{URL }\fi
\providecommand{\bibinfo}[2]{#2}
\providecommand{\eprint}[2][]{\url{#2}}

\bibitem[{\citenamefont{Jaksch et~al.}(1998)\citenamefont{Jaksch, Bruder,
  Cirac, Gardiner, and Zoller}}]{jaksch_cold_1998}
\bibinfo{author}{\bibfnamefont{D.}~\bibnamefont{Jaksch}},
  \bibinfo{author}{\bibfnamefont{C.}~\bibnamefont{Bruder}},
  \bibinfo{author}{\bibfnamefont{J.~I.} \bibnamefont{Cirac}},
  \bibinfo{author}{\bibfnamefont{C.~W.} \bibnamefont{Gardiner}},
  \bibnamefont{and} \bibinfo{author}{\bibfnamefont{P.}~\bibnamefont{Zoller}},
  \bibinfo{journal}{Phys. Rev. Lett.} \textbf{\bibinfo{volume}{81}},
  \bibinfo{pages}{3108} (\bibinfo{year}{1998}).

\bibitem[{\citenamefont{Greiner et~al.}(2002)\citenamefont{Greiner, Mandel,
  Esslinger, H{\"a}nsch, and Bloch}}]{greiner_quantum_2002}
\bibinfo{author}{\bibfnamefont{M.}~\bibnamefont{Greiner}},
  \bibinfo{author}{\bibfnamefont{O.}~\bibnamefont{Mandel}},
  \bibinfo{author}{\bibfnamefont{T.}~\bibnamefont{Esslinger}},
  \bibinfo{author}{\bibfnamefont{T.~W.} \bibnamefont{H{\"a}nsch}},
  \bibnamefont{and} \bibinfo{author}{\bibfnamefont{I.}~\bibnamefont{Bloch}},
  \bibinfo{journal}{Nature (London)} \textbf{\bibinfo{volume}{415}},
  \bibinfo{pages}{39} (\bibinfo{year}{2002}).

\bibitem[{\citenamefont{Bloch et~al.}(2008)\citenamefont{Bloch, Dalibard, and
  Zwerger}}]{bloch_many-body_2008}
\bibinfo{author}{\bibfnamefont{I.}~\bibnamefont{Bloch}},
  \bibinfo{author}{\bibfnamefont{J.}~\bibnamefont{Dalibard}}, \bibnamefont{and}
  \bibinfo{author}{\bibfnamefont{W.}~\bibnamefont{Zwerger}},
  \bibinfo{journal}{Rev. Mod. Phys.} \textbf{\bibinfo{volume}{80}},
  \bibinfo{pages}{885} (\bibinfo{year}{2008}).

\bibitem[{\citenamefont{Hartmann et~al.}(2008)\citenamefont{Hartmann,
  Brand{\~a}o, and Plenio}}]{hartmann_quantum_2008}
\bibinfo{author}{\bibfnamefont{M.}~\bibnamefont{Hartmann}},
  \bibinfo{author}{\bibfnamefont{F.~G.} \bibnamefont{Brand{\~a}o}},
  \bibnamefont{and} \bibinfo{author}{\bibfnamefont{M.~B.}
  \bibnamefont{Plenio}}, \bibinfo{journal}{Laser \& Photonics Rev.}
  \textbf{\bibinfo{volume}{2}}, \bibinfo{pages}{527} (\bibinfo{year}{2008}).

\bibitem[{\citenamefont{Tomadin and Fazio}(2010)}]{to.fa.10}
\bibinfo{author}{\bibfnamefont{A.}~\bibnamefont{Tomadin}} \bibnamefont{and}
  \bibinfo{author}{\bibfnamefont{R.}~\bibnamefont{Fazio}}, \bibinfo{journal}{J.
  Opt. Soc. Am. B} \textbf{\bibinfo{volume}{27}}, \bibinfo{pages}{A130}
  (\bibinfo{year}{2010}).

\bibitem[{\citenamefont{Park et~al.}(2000)\citenamefont{Park, Park, Lim,
  Anderson, Alivisatos, and McEuen}}]{pa.pa.00}
\bibinfo{author}{\bibfnamefont{H.}~\bibnamefont{Park}},
  \bibinfo{author}{\bibfnamefont{J.}~\bibnamefont{Park}},
  \bibinfo{author}{\bibfnamefont{A.~K.~L.} \bibnamefont{Lim}},
  \bibinfo{author}{\bibfnamefont{E.~H.} \bibnamefont{Anderson}},
  \bibinfo{author}{\bibfnamefont{A.~P.} \bibnamefont{Alivisatos}},
  \bibnamefont{and} \bibinfo{author}{\bibfnamefont{P.~L.}
  \bibnamefont{McEuen}}, \bibinfo{journal}{Nature}
  \textbf{\bibinfo{volume}{407}}, \bibinfo{pages}{57} (\bibinfo{year}{2000}).

\bibitem[{\citenamefont{Paaske and Flensberg}(2005)}]{pa.fl.05}
\bibinfo{author}{\bibfnamefont{J.}~\bibnamefont{Paaske}} \bibnamefont{and}
  \bibinfo{author}{\bibfnamefont{K.}~\bibnamefont{Flensberg}},
  \bibinfo{journal}{Phys. Rev. Lett.} \textbf{\bibinfo{volume}{94}},
  \bibinfo{pages}{176801} (\bibinfo{year}{2005}).

\bibitem[{\citenamefont{Zutic et~al.}(2004)\citenamefont{Zutic, Fabian, and
  Sarma}}]{zu.fa.04}
\bibinfo{author}{\bibfnamefont{I.}~\bibnamefont{Zutic}},
  \bibinfo{author}{\bibfnamefont{J.}~\bibnamefont{Fabian}}, \bibnamefont{and}
  \bibinfo{author}{\bibfnamefont{S.~D.} \bibnamefont{Sarma}},
  \bibinfo{journal}{Rev. Mod. Phys.} \textbf{\bibinfo{volume}{76}},
  \bibinfo{pages}{323} (\bibinfo{year}{2004}).

\bibitem[{\citenamefont{Fabian et~al.}(2007)\citenamefont{Fabian,
  Matos-Abiague, Ertler, Stano, and Zutic}}]{fa.ma.07}
\bibinfo{author}{\bibfnamefont{J.}~\bibnamefont{Fabian}},
  \bibinfo{author}{\bibfnamefont{A.}~\bibnamefont{Matos-Abiague}},
  \bibinfo{author}{\bibfnamefont{C.}~\bibnamefont{Ertler}},
  \bibinfo{author}{\bibfnamefont{P.}~\bibnamefont{Stano}}, \bibnamefont{and}
  \bibinfo{author}{\bibfnamefont{I.}~\bibnamefont{Zutic}},
  \bibinfo{journal}{Acta Physica Slovaca} \textbf{\bibinfo{volume}{57}},
  \bibinfo{pages}{565} (\bibinfo{year}{2007}).

\bibitem[{\citenamefont{Slobodskyy et~al.}(2003)\citenamefont{Slobodskyy,
  Gould, Slobodskyy, Becker, Schmidt, and Molenkamp}}]{sl.go.03}
\bibinfo{author}{\bibfnamefont{A.}~\bibnamefont{Slobodskyy}},
  \bibinfo{author}{\bibfnamefont{C.}~\bibnamefont{Gould}},
  \bibinfo{author}{\bibfnamefont{T.}~\bibnamefont{Slobodskyy}},
  \bibinfo{author}{\bibfnamefont{C.~R.} \bibnamefont{Becker}},
  \bibinfo{author}{\bibfnamefont{G.}~\bibnamefont{Schmidt}}, \bibnamefont{and}
  \bibinfo{author}{\bibfnamefont{L.~W.} \bibnamefont{Molenkamp}},
  \bibinfo{journal}{Phys. Rev. Lett.} \textbf{\bibinfo{volume}{90}},
  \bibinfo{pages}{246601} (\bibinfo{year}{2003}).

\bibitem[{\citenamefont{Bonilla and Grahn}(2005)}]{bonilla_non-linear_2005}
\bibinfo{author}{\bibfnamefont{L.~L.} \bibnamefont{Bonilla}} \bibnamefont{and}
  \bibinfo{author}{\bibfnamefont{H.~T.} \bibnamefont{Grahn}},
  \bibinfo{journal}{Rep. Prog. Phys.} \textbf{\bibinfo{volume}{68}},
  \bibinfo{pages}{577} (\bibinfo{year}{2005}).

\bibitem[{\citenamefont{Jungwirth et~al.}(2006)\citenamefont{Jungwirth, Sinova,
  Masek, Kucera, and {MacDonald}}}]{jungwirth_theory_2006}
\bibinfo{author}{\bibfnamefont{T.}~\bibnamefont{Jungwirth}},
  \bibinfo{author}{\bibfnamefont{J.}~\bibnamefont{Sinova}},
  \bibinfo{author}{\bibfnamefont{J.}~\bibnamefont{Masek}},
  \bibinfo{author}{\bibfnamefont{J.}~\bibnamefont{Kucera}}, \bibnamefont{and}
  \bibinfo{author}{\bibfnamefont{A.~H.} \bibnamefont{{MacDonald}}},
  \bibinfo{journal}{Rev. Mod. Phys.} \textbf{\bibinfo{volume}{78}},
  \bibinfo{pages}{809} (\bibinfo{year}{2006}).

\bibitem[{\citenamefont{Ertler et~al.}(2010{\natexlab{a}})\citenamefont{Ertler,
  P{\"o}tz, and Fabian}}]{ertler_proposal_2010}
\bibinfo{author}{\bibfnamefont{C.}~\bibnamefont{Ertler}},
  \bibinfo{author}{\bibfnamefont{W.}~\bibnamefont{P{\"o}tz}}, \bibnamefont{and}
  \bibinfo{author}{\bibfnamefont{J.}~\bibnamefont{Fabian}},
  \bibinfo{journal}{Appl. Phys. Lett.} \textbf{\bibinfo{volume}{97}},
  \bibinfo{pages}{042104} (\bibinfo{year}{2010}{\natexlab{a}}).

\bibitem[{\citenamefont{Ertler et~al.}(2010{\natexlab{b}})\citenamefont{Ertler,
  Senekowitsch, Fabian, and P{\"o}tz}}]{ertler_self-consistent}
\bibinfo{author}{\bibfnamefont{C.}~\bibnamefont{Ertler}},
  \bibinfo{author}{\bibfnamefont{P.}~\bibnamefont{Senekowitsch}},
  \bibinfo{author}{\bibfnamefont{J.}~\bibnamefont{Fabian}}, \bibnamefont{and}
  \bibinfo{author}{\bibfnamefont{W.}~\bibnamefont{P{\"o}tz}}, in
  \emph{\bibinfo{booktitle}{Computational Electronics (IWCE), 2010 14th
  International Workshop on}} (\bibinfo{year}{2010}{\natexlab{b}}), pp.
  \bibinfo{pages}{1--4}.

\bibitem[{\citenamefont{Sokolowski-Tinten
  et~al.}(2003)\citenamefont{Sokolowski-Tinten, Blome, Blums, Cavalleri,
  Dietrich, Tarasevitch, Uschmann, Forster, Kammler, Horn-von Hoegen
  et~al.}}]{so.bl.03}
\bibinfo{author}{\bibfnamefont{K.}~\bibnamefont{Sokolowski-Tinten}},
  \bibinfo{author}{\bibfnamefont{C.}~\bibnamefont{Blome}},
  \bibinfo{author}{\bibfnamefont{J.}~\bibnamefont{Blums}},
  \bibinfo{author}{\bibfnamefont{A.}~\bibnamefont{Cavalleri}},
  \bibinfo{author}{\bibfnamefont{C.}~\bibnamefont{Dietrich}},
  \bibinfo{author}{\bibfnamefont{A.}~\bibnamefont{Tarasevitch}},
  \bibinfo{author}{\bibfnamefont{I.}~\bibnamefont{Uschmann}},
  \bibinfo{author}{\bibfnamefont{E.}~\bibnamefont{Forster}},
  \bibinfo{author}{\bibfnamefont{M.}~\bibnamefont{Kammler}},
  \bibinfo{author}{\bibfnamefont{M.}~\bibnamefont{Horn-von Hoegen}},
  \bibnamefont{et~al.}, \bibinfo{journal}{Nature}
  \textbf{\bibinfo{volume}{422}}, \bibinfo{pages}{287} (\bibinfo{year}{2003}).

\bibitem[{\citenamefont{Haug and Jauho}(1998)}]{ha.ja}
\bibinfo{author}{\bibfnamefont{H.}~\bibnamefont{Haug}} \bibnamefont{and}
  \bibinfo{author}{\bibfnamefont{A.-P.} \bibnamefont{Jauho}},
  \emph{\bibinfo{title}{Quantum Kinetics in Transport and Optics of
  Semiconductors}} (\bibinfo{publisher}{Springer},
  \bibinfo{address}{Heidelberg}, \bibinfo{year}{1998}).

\bibitem[{\citenamefont{Rammer and Smith}(1986)}]{ra.sm.86}
\bibinfo{author}{\bibfnamefont{J.}~\bibnamefont{Rammer}} \bibnamefont{and}
  \bibinfo{author}{\bibfnamefont{H.}~\bibnamefont{Smith}},
  \bibinfo{journal}{Rev. Mod. Phys.} \textbf{\bibinfo{volume}{58}},
  \bibinfo{pages}{323} (\bibinfo{year}{1986}).

\bibitem[{\citenamefont{Meir and Wingreen}(1992)}]{me.wi.92}
\bibinfo{author}{\bibfnamefont{Y.}~\bibnamefont{Meir}} \bibnamefont{and}
  \bibinfo{author}{\bibfnamefont{N.~S.} \bibnamefont{Wingreen}},
  \bibinfo{journal}{Phys. Rev. Lett.} \textbf{\bibinfo{volume}{68}},
  \bibinfo{pages}{2512} (\bibinfo{year}{1992}).

\bibitem[{\citenamefont{Meir et~al.}(1993)\citenamefont{Meir, Wingreen, and
  Lee}}]{me.wi.93}
\bibinfo{author}{\bibfnamefont{Y.}~\bibnamefont{Meir}},
  \bibinfo{author}{\bibfnamefont{N.~S.} \bibnamefont{Wingreen}},
  \bibnamefont{and} \bibinfo{author}{\bibfnamefont{P.~A.} \bibnamefont{Lee}},
  \bibinfo{journal}{Phys. Rev. Lett.} \textbf{\bibinfo{volume}{70}},
  \bibinfo{pages}{2601} (\bibinfo{year}{1993}).

\bibitem[{\citenamefont{Ryndyk et~al.}(2009)\citenamefont{Ryndyk, Gutierrez,
  Song, and Cuniberti}}]{ry.gu.09}
\bibinfo{author}{\bibfnamefont{D.~A.} \bibnamefont{Ryndyk}},
  \bibinfo{author}{\bibfnamefont{R.}~\bibnamefont{Gutierrez}},
  \bibinfo{author}{\bibfnamefont{B.}~\bibnamefont{Song}}, \bibnamefont{and}
  \bibinfo{author}{\bibfnamefont{G.}~\bibnamefont{Cuniberti}}, in
  \emph{\bibinfo{booktitle}{Energy Transfer Dynamics in Biomaterial Systems}},
  edited by \bibinfo{editor}{\bibfnamefont{A.~W.} \bibnamefont{Castleman}},
  \bibinfo{editor}{\bibfnamefont{J.~P.} \bibnamefont{Toennies}},
  \bibinfo{editor}{\bibfnamefont{K.}~\bibnamefont{Yamanouchi}},
  \bibinfo{editor}{\bibfnamefont{W.}~\bibnamefont{Zinth}},
  \bibinfo{editor}{\bibfnamefont{I.}~\bibnamefont{Burghardt}},
  \bibinfo{editor}{\bibfnamefont{V.}~\bibnamefont{May}},
  \bibinfo{editor}{\bibfnamefont{D.~A.} \bibnamefont{Micha}}, \bibnamefont{and}
  \bibinfo{editor}{\bibfnamefont{E.~R.} \bibnamefont{Bittner}}
  (\bibinfo{publisher}{Springer Berlin Heidelberg}, \bibinfo{year}{2009}),
  vol.~\bibinfo{volume}{93} of \emph{\bibinfo{series}{Springer Series in
  Chemical Physics}}, pp. \bibinfo{pages}{213--335}.

\bibitem[{\citenamefont{{Schoeller, H.}}(2009)}]{scho.09}
\bibinfo{author}{\bibnamefont{{Schoeller, H.}}}, \bibinfo{journal}{Eur. Phys.
  J. Special Topics} \textbf{\bibinfo{volume}{168}}, \bibinfo{pages}{179}
  (\bibinfo{year}{2009}).

\bibitem[{\citenamefont{Diehl et~al.}(2008)\citenamefont{Diehl, Micheli,
  Kantian, Kraus, B{\"u}chler, and Zoller}}]{di.mi.08}
\bibinfo{author}{\bibfnamefont{S.}~\bibnamefont{Diehl}},
  \bibinfo{author}{\bibfnamefont{A.}~\bibnamefont{Micheli}},
  \bibinfo{author}{\bibfnamefont{A.}~\bibnamefont{Kantian}},
  \bibinfo{author}{\bibfnamefont{B.}~\bibnamefont{Kraus}},
  \bibinfo{author}{\bibfnamefont{H.~P.} \bibnamefont{B{\"u}chler}},
  \bibnamefont{and} \bibinfo{author}{\bibfnamefont{P.}~\bibnamefont{Zoller}},
  \bibinfo{journal}{Nat. Phys.} \textbf{\bibinfo{volume}{4}},
  \bibinfo{pages}{878 } (\bibinfo{year}{2008}).

\bibitem[{\citenamefont{Kraus et~al.}(2008)\citenamefont{Kraus, B\"uchler,
  Diehl, Kantian, Micheli, and Zoller}}]{kr.bu.08}
\bibinfo{author}{\bibfnamefont{B.}~\bibnamefont{Kraus}},
  \bibinfo{author}{\bibfnamefont{H.~P.} \bibnamefont{B\"uchler}},
  \bibinfo{author}{\bibfnamefont{S.}~\bibnamefont{Diehl}},
  \bibinfo{author}{\bibfnamefont{A.}~\bibnamefont{Kantian}},
  \bibinfo{author}{\bibfnamefont{A.}~\bibnamefont{Micheli}}, \bibnamefont{and}
  \bibinfo{author}{\bibfnamefont{P.}~\bibnamefont{Zoller}},
  \bibinfo{journal}{Phys. Rev. A} \textbf{\bibinfo{volume}{78}},
  \bibinfo{pages}{042307} (\bibinfo{year}{2008}).

\bibitem[{\citenamefont{Diehl et~al.}(2010)\citenamefont{Diehl, Tomadin,
  Micheli, Fazio, and Zoller}}]{di.to.10}
\bibinfo{author}{\bibfnamefont{S.}~\bibnamefont{Diehl}},
  \bibinfo{author}{\bibfnamefont{A.}~\bibnamefont{Tomadin}},
  \bibinfo{author}{\bibfnamefont{A.}~\bibnamefont{Micheli}},
  \bibinfo{author}{\bibfnamefont{R.}~\bibnamefont{Fazio}}, \bibnamefont{and}
  \bibinfo{author}{\bibfnamefont{P.}~\bibnamefont{Zoller}},
  \bibinfo{journal}{Phys. Rev. Lett.} \textbf{\bibinfo{volume}{105}},
  \bibinfo{pages}{015702} (\bibinfo{year}{2010}).

\bibitem[{\citenamefont{Pichler et~al.}(2010)\citenamefont{Pichler, Daley, and
  Zoller}}]{pi.da.10}
\bibinfo{author}{\bibfnamefont{H.}~\bibnamefont{Pichler}},
  \bibinfo{author}{\bibfnamefont{A.~J.} \bibnamefont{Daley}}, \bibnamefont{and}
  \bibinfo{author}{\bibfnamefont{P.}~\bibnamefont{Zoller}},
  \bibinfo{journal}{Phys. Rev. A} \textbf{\bibinfo{volume}{82}},
  \bibinfo{pages}{063605} (\bibinfo{year}{2010}).

\bibitem[{\citenamefont{Tomadin et~al.}(2011)\citenamefont{Tomadin, Diehl, and
  Zoller}}]{tomadin_nonequilibrium_2011}
\bibinfo{author}{\bibfnamefont{A.}~\bibnamefont{Tomadin}},
  \bibinfo{author}{\bibfnamefont{S.}~\bibnamefont{Diehl}}, \bibnamefont{and}
  \bibinfo{author}{\bibfnamefont{P.}~\bibnamefont{Zoller}},
  \bibinfo{journal}{Phys. Rev. A} \textbf{\bibinfo{volume}{83}},
  \bibinfo{pages}{013611} (\bibinfo{year}{2011}).

\bibitem[{\citenamefont{Barmettler and
  Kollath}(2010)}]{barmettler_dynamical_2010}
\bibinfo{author}{\bibfnamefont{P.}~\bibnamefont{Barmettler}} \bibnamefont{and}
  \bibinfo{author}{\bibfnamefont{C.}~\bibnamefont{Kollath}},
  \bibinfo{journal}{arXiv:1012.0422}  (\bibinfo{year}{2010}).

\bibitem[{\citenamefont{{Dalla Torre} et~al.}(2010)\citenamefont{{Dalla Torre},
  Demler, Giamarchi, and Altman}}]{dalla_torre_quantum_2010}
\bibinfo{author}{\bibfnamefont{E.~G.} \bibnamefont{{Dalla Torre}}},
  \bibinfo{author}{\bibfnamefont{E.}~\bibnamefont{Demler}},
  \bibinfo{author}{\bibfnamefont{T.}~\bibnamefont{Giamarchi}},
  \bibnamefont{and} \bibinfo{author}{\bibfnamefont{E.}~\bibnamefont{Altman}},
  \bibinfo{journal}{Nat. Phys.} \textbf{\bibinfo{volume}{6}},
  \bibinfo{pages}{806} (\bibinfo{year}{2010}).

\bibitem[{\citenamefont{White and Feiguin}(2004)}]{wh.fe.04}
\bibinfo{author}{\bibfnamefont{S.~R.} \bibnamefont{White}} \bibnamefont{and}
  \bibinfo{author}{\bibfnamefont{A.~E.} \bibnamefont{Feiguin}},
  \bibinfo{journal}{Phys. Rev. Lett.} \textbf{\bibinfo{volume}{93}},
  \bibinfo{pages}{076401} (\bibinfo{year}{2004}).

\bibitem[{\citenamefont{Daley et~al.}(2004)\citenamefont{Daley, Kollath,
  Schollw{\"o}ck, and Vidal}}]{daley_time-dependent_2004}
\bibinfo{author}{\bibfnamefont{A.~J.} \bibnamefont{Daley}},
  \bibinfo{author}{\bibfnamefont{C.}~\bibnamefont{Kollath}},
  \bibinfo{author}{\bibfnamefont{U.}~\bibnamefont{Schollw{\"o}ck}},
  \bibnamefont{and} \bibinfo{author}{\bibfnamefont{G.}~\bibnamefont{Vidal}},
  \bibinfo{journal}{J. Stat. Mech.} \textbf{\bibinfo{volume}{2004}},
  \bibinfo{pages}{P04005} (\bibinfo{year}{2004}).

\bibitem[{\citenamefont{Prosen and {\v Z}nidari{\v
  c}}(2009)}]{prosen_matrix_2009}
\bibinfo{author}{\bibfnamefont{T.}~\bibnamefont{Prosen}} \bibnamefont{and}
  \bibinfo{author}{\bibfnamefont{M.}~\bibnamefont{{\v Z}nidari{\v c}}},
  \bibinfo{journal}{J. Stat. Mech.} \textbf{\bibinfo{volume}{2009}},
  \bibinfo{pages}{P02035} (\bibinfo{year}{2009}).

\bibitem[{\citenamefont{Benenti et~al.}(2009)\citenamefont{Benenti, Casati,
  Prosen, Rossini, and {\v{Z}}nidari{\v{c}}}}]{be.ca.09}
\bibinfo{author}{\bibfnamefont{G.}~\bibnamefont{Benenti}},
  \bibinfo{author}{\bibfnamefont{G.}~\bibnamefont{Casati}},
  \bibinfo{author}{\bibfnamefont{T.}~\bibnamefont{Prosen}},
  \bibinfo{author}{\bibfnamefont{D.}~\bibnamefont{Rossini}}, \bibnamefont{and}
  \bibinfo{author}{\bibfnamefont{M.}~\bibnamefont{{\v{Z}}nidari{\v{c}}}},
  \bibinfo{journal}{Phys. Rev. B} \textbf{\bibinfo{volume}{80}},
  \bibinfo{pages}{035110} (\bibinfo{year}{2009}).

\bibitem[{\citenamefont{Perez-Garcia et~al.}(2007)\citenamefont{Perez-Garcia,
  Verstraete, Wolf, and Cirac}}]{pe.ve.07}
\bibinfo{author}{\bibfnamefont{D.}~\bibnamefont{Perez-Garcia}},
  \bibinfo{author}{\bibfnamefont{F.}~\bibnamefont{Verstraete}},
  \bibinfo{author}{\bibfnamefont{M.~M.} \bibnamefont{Wolf}}, \bibnamefont{and}
  \bibinfo{author}{\bibfnamefont{J.~I.} \bibnamefont{Cirac}},
  \bibinfo{journal}{Quant. Inf. Comp.} \textbf{\bibinfo{volume}{7}},
  \bibinfo{pages}{401} (\bibinfo{year}{2007}).

\bibitem[{\citenamefont{Werner et~al.}(2009)\citenamefont{Werner, Oka, and
  Millis}}]{we.ok.09}
\bibinfo{author}{\bibfnamefont{P.}~\bibnamefont{Werner}},
  \bibinfo{author}{\bibfnamefont{T.}~\bibnamefont{Oka}}, \bibnamefont{and}
  \bibinfo{author}{\bibfnamefont{A.~J.} \bibnamefont{Millis}},
  \bibinfo{journal}{Phys. Rev. B} \textbf{\bibinfo{volume}{79}},
  \bibinfo{pages}{035320} (\bibinfo{year}{2009}).

\bibitem[{\citenamefont{Anders and Schiller}(2006)}]{an.sc.06}
\bibinfo{author}{\bibfnamefont{F.~B.} \bibnamefont{Anders}} \bibnamefont{and}
  \bibinfo{author}{\bibfnamefont{A.}~\bibnamefont{Schiller}},
  \bibinfo{journal}{Phys. Rev. B} \textbf{\bibinfo{volume}{74}},
  \bibinfo{pages}{245113} (\bibinfo{year}{2006}).

\bibitem[{\citenamefont{Jakobs et~al.}(2007)\citenamefont{Jakobs, Meden, and
  Schoeller}}]{ja.me.07}
\bibinfo{author}{\bibfnamefont{S.~G.} \bibnamefont{Jakobs}},
  \bibinfo{author}{\bibfnamefont{V.}~\bibnamefont{Meden}}, \bibnamefont{and}
  \bibinfo{author}{\bibfnamefont{H.}~\bibnamefont{Schoeller}},
  \bibinfo{journal}{Phys. Rev. Lett.} \textbf{\bibinfo{volume}{99}},
  \bibinfo{pages}{150603} (\bibinfo{year}{2007}).

\bibitem[{\citenamefont{Joura et~al.}(2008)\citenamefont{Joura, Freericks, and
  Pruschke}}]{jo.fr.08}
\bibinfo{author}{\bibfnamefont{A.~V.} \bibnamefont{Joura}},
  \bibinfo{author}{\bibfnamefont{J.~K.} \bibnamefont{Freericks}},
  \bibnamefont{and} \bibinfo{author}{\bibfnamefont{T.}~\bibnamefont{Pruschke}},
  \bibinfo{journal}{Phys. Rev. Lett.} \textbf{\bibinfo{volume}{101}},
  \bibinfo{pages}{196401} (\bibinfo{year}{2008}).

\bibitem[{\citenamefont{Eckstein et~al.}(2009)\citenamefont{Eckstein, Kollar,
  and Werner}}]{eckstein_thermalization_2009}
\bibinfo{author}{\bibfnamefont{M.}~\bibnamefont{Eckstein}},
  \bibinfo{author}{\bibfnamefont{M.}~\bibnamefont{Kollar}}, \bibnamefont{and}
  \bibinfo{author}{\bibfnamefont{P.}~\bibnamefont{Werner}},
  \bibinfo{journal}{Phys. Rev. Lett.} \textbf{\bibinfo{volume}{103}},
  \bibinfo{pages}{056403} (\bibinfo{year}{2009}).

\bibitem[{\citenamefont{Aron et~al.}(2011)\citenamefont{Aron, Kotliar, and
  Weber}}]{ar.ko.11u}
\bibinfo{author}{\bibfnamefont{C.}~\bibnamefont{Aron}},
  \bibinfo{author}{\bibfnamefont{G.}~\bibnamefont{Kotliar}}, \bibnamefont{and}
  \bibinfo{author}{\bibfnamefont{C.}~\bibnamefont{Weber}}
  (\bibinfo{year}{2011}), \bibinfo{note}{arXiv:1105.5387}.

\bibitem[{\citenamefont{Freericks et~al.}(2006)\citenamefont{Freericks,
  Turkowski, and Zlati{\'{c}}}}]{fr.tu.06}
\bibinfo{author}{\bibfnamefont{J.~K.} \bibnamefont{Freericks}},
  \bibinfo{author}{\bibfnamefont{V.~M.} \bibnamefont{Turkowski}},
  \bibnamefont{and}
  \bibinfo{author}{\bibfnamefont{V.}~\bibnamefont{Zlati{\'{c}}}},
  \bibinfo{journal}{Phys. Rev. Lett.} \textbf{\bibinfo{volume}{97}},
  \bibinfo{pages}{266408} (\bibinfo{year}{2006}).

\bibitem[{\citenamefont{Mehta and Andrei}(2006)}]{me.an.06}
\bibinfo{author}{\bibfnamefont{P.}~\bibnamefont{Mehta}} \bibnamefont{and}
  \bibinfo{author}{\bibfnamefont{N.}~\bibnamefont{Andrei}},
  \bibinfo{journal}{Phys. Rev. Lett.} \textbf{\bibinfo{volume}{96}},
  \bibinfo{pages}{216802} (\bibinfo{year}{2006}).

\bibitem[{\citenamefont{Gritsev et~al.}(2010)\citenamefont{Gritsev, Rostunov,
  and Demler}}]{gritsev_exact_2010}
\bibinfo{author}{\bibfnamefont{V.}~\bibnamefont{Gritsev}},
  \bibinfo{author}{\bibfnamefont{T.}~\bibnamefont{Rostunov}}, \bibnamefont{and}
  \bibinfo{author}{\bibfnamefont{E.}~\bibnamefont{Demler}},
  \bibinfo{journal}{J. Stat. Mech.} \textbf{\bibinfo{volume}{2010}},
  \bibinfo{pages}{P05012} (\bibinfo{year}{2010}).

\bibitem[{\citenamefont{Jung et~al.}(2010)\citenamefont{Jung, Lieder, Brener,
  Hafermann, Baxevanis, Chudnovskiy, Rubtsov, Katsnelson, and
  Lichtenstein}}]{jung_dual-fermion_2010}
\bibinfo{author}{\bibfnamefont{C.}~\bibnamefont{Jung}},
  \bibinfo{author}{\bibfnamefont{A.}~\bibnamefont{Lieder}},
  \bibinfo{author}{\bibfnamefont{S.}~\bibnamefont{Brener}},
  \bibinfo{author}{\bibfnamefont{H.}~\bibnamefont{Hafermann}},
  \bibinfo{author}{\bibfnamefont{B.}~\bibnamefont{Baxevanis}},
  \bibinfo{author}{\bibfnamefont{A.}~\bibnamefont{Chudnovskiy}},
  \bibinfo{author}{\bibfnamefont{A.~N.} \bibnamefont{Rubtsov}},
  \bibinfo{author}{\bibfnamefont{M.~I.} \bibnamefont{Katsnelson}},
  \bibnamefont{and} \bibinfo{author}{\bibfnamefont{A.~I.}
  \bibnamefont{Lichtenstein}}, \bibinfo{journal}{arXiv:1011.3264}
  (\bibinfo{year}{2010}).

\bibitem[{\citenamefont{Balzer and
  Potthoff}(2011)}]{balzer_nonequilibrium_2011}
\bibinfo{author}{\bibfnamefont{M.}~\bibnamefont{Balzer}} \bibnamefont{and}
  \bibinfo{author}{\bibfnamefont{M.}~\bibnamefont{Potthoff}},
  \bibinfo{journal}{Phys. Rev. B} \textbf{\bibinfo{volume}{83}},
  \bibinfo{pages}{195132} (\bibinfo{year}{2011}).

\bibitem[{\citenamefont{My{\"o}h{\"a}nen
  et~al.}(2009)\citenamefont{My{\"o}h{\"a}nen, Stan, Stefanucci, and van
  Leeuwen}}]{my.st.09}
\bibinfo{author}{\bibfnamefont{P.}~\bibnamefont{My{\"o}h{\"a}nen}},
  \bibinfo{author}{\bibfnamefont{A.}~\bibnamefont{Stan}},
  \bibinfo{author}{\bibfnamefont{G.}~\bibnamefont{Stefanucci}},
  \bibnamefont{and} \bibinfo{author}{\bibfnamefont{R.}~\bibnamefont{van
  Leeuwen}}, \bibinfo{journal}{Phys. Rev. B} \textbf{\bibinfo{volume}{80}},
  \bibinfo{pages}{115107} (\bibinfo{year}{2009}).

\bibitem[{\citenamefont{Brandbyge et~al.}(2002)\citenamefont{Brandbyge, Mozos,
  Ordej{\'o}n, Taylor, and Stokbro}}]{Brandbyge:2002}
\bibinfo{author}{\bibfnamefont{M.}~\bibnamefont{Brandbyge}},
  \bibinfo{author}{\bibfnamefont{J.-L.} \bibnamefont{Mozos}},
  \bibinfo{author}{\bibfnamefont{P.}~\bibnamefont{Ordej{\'o}n}},
  \bibinfo{author}{\bibfnamefont{J.}~\bibnamefont{Taylor}}, \bibnamefont{and}
  \bibinfo{author}{\bibfnamefont{K.}~\bibnamefont{Stokbro}},
  \bibinfo{journal}{Phys. Rev. B} \textbf{\bibinfo{volume}{65}},
  \bibinfo{pages}{165401} (\bibinfo{year}{2002}).

\bibitem[{\citenamefont{F\"urst et~al.}(2008)\citenamefont{F\"urst, Brandbyge,
  Jauho, and Stokbro}}]{fu.br.08}
\bibinfo{author}{\bibfnamefont{J.~A.} \bibnamefont{F\"urst}},
  \bibinfo{author}{\bibfnamefont{M.}~\bibnamefont{Brandbyge}},
  \bibinfo{author}{\bibfnamefont{A.-P.} \bibnamefont{Jauho}}, \bibnamefont{and}
  \bibinfo{author}{\bibfnamefont{K.}~\bibnamefont{Stokbro}},
  \bibinfo{journal}{Phys. Rev. B} \textbf{\bibinfo{volume}{78}},
  \bibinfo{pages}{195405} (\bibinfo{year}{2008}).

\bibitem[{\citenamefont{Markussen et~al.}(2009)\citenamefont{Markussen, Jauho,
  and Brandbyge}}]{ma.ja.09}
\bibinfo{author}{\bibfnamefont{T.}~\bibnamefont{Markussen}},
  \bibinfo{author}{\bibfnamefont{A.-P.} \bibnamefont{Jauho}}, \bibnamefont{and}
  \bibinfo{author}{\bibfnamefont{M.}~\bibnamefont{Brandbyge}},
  \bibinfo{journal}{Phys. Rev. B} \textbf{\bibinfo{volume}{79}},
  \bibinfo{pages}{035415} (\bibinfo{year}{2009}).

\bibitem[{\citenamefont{Dahnken et~al.}(2004)\citenamefont{Dahnken, Aichhorn,
  Hanke, Arrigoni, and Potthoff}}]{da.ai.04}
\bibinfo{author}{\bibfnamefont{C.}~\bibnamefont{Dahnken}},
  \bibinfo{author}{\bibfnamefont{M.}~\bibnamefont{Aichhorn}},
  \bibinfo{author}{\bibfnamefont{W.}~\bibnamefont{Hanke}},
  \bibinfo{author}{\bibfnamefont{E.}~\bibnamefont{Arrigoni}}, \bibnamefont{and}
  \bibinfo{author}{\bibfnamefont{M.}~\bibnamefont{Potthoff}},
  \bibinfo{journal}{Phys. Rev. B} \textbf{\bibinfo{volume}{70}},
  \bibinfo{pages}{245110} (\bibinfo{year}{2004}).

\bibitem[{\citenamefont{Knap et~al.}(2011)\citenamefont{Knap, Arrigoni, and
  von~der Linden}}]{kn.ar.11}
\bibinfo{author}{\bibfnamefont{M.}~\bibnamefont{Knap}},
  \bibinfo{author}{\bibfnamefont{E.}~\bibnamefont{Arrigoni}}, \bibnamefont{and}
  \bibinfo{author}{\bibfnamefont{W.}~\bibnamefont{von~der Linden}},
  \bibinfo{journal}{Phys. Rev. B} \textbf{\bibinfo{volume}{83}},
  \bibinfo{pages}{134507} (\bibinfo{year}{2011}).

\bibitem[{\citenamefont{Metzner and Vollhardt}(1989)}]{me.vo.89}
\bibinfo{author}{\bibfnamefont{W.}~\bibnamefont{Metzner}} \bibnamefont{and}
  \bibinfo{author}{\bibfnamefont{D.}~\bibnamefont{Vollhardt}},
  \bibinfo{journal}{Phys. Rev. Lett.} \textbf{\bibinfo{volume}{62}},
  \bibinfo{pages}{324} (\bibinfo{year}{1989}).

\bibitem[{\citenamefont{Georges et~al.}(1996)\citenamefont{Georges, Kotliar,
  Krauth, and Rozenberg}}]{ge.ko.96}
\bibinfo{author}{\bibfnamefont{A.}~\bibnamefont{Georges}},
  \bibinfo{author}{\bibfnamefont{G.}~\bibnamefont{Kotliar}},
  \bibinfo{author}{\bibfnamefont{W.}~\bibnamefont{Krauth}}, \bibnamefont{and}
  \bibinfo{author}{\bibfnamefont{M.~J.} \bibnamefont{Rozenberg}},
  \bibinfo{journal}{Rev. Mod. Phys.} \textbf{\bibinfo{volume}{68}},
  \bibinfo{pages}{13} (\bibinfo{year}{1996}).

\bibitem[{\citenamefont{Kotliar et~al.}(2001)\citenamefont{Kotliar, Savrasov,
  P{\'a}lsson, and Biroli}}]{ko.sa.01}
\bibinfo{author}{\bibfnamefont{G.}~\bibnamefont{Kotliar}},
  \bibinfo{author}{\bibfnamefont{S.~Y.} \bibnamefont{Savrasov}},
  \bibinfo{author}{\bibfnamefont{G.}~\bibnamefont{P{\'a}lsson}},
  \bibnamefont{and} \bibinfo{author}{\bibfnamefont{G.}~\bibnamefont{Biroli}},
  \bibinfo{journal}{Phys. Rev. Lett.} \textbf{\bibinfo{volume}{87}},
  \bibinfo{pages}{186401} (\bibinfo{year}{2001}).

\bibitem[{\citenamefont{Potthoff}(2003{\natexlab{a}})}]{pott.03}
\bibinfo{author}{\bibfnamefont{M.}~\bibnamefont{Potthoff}},
  \bibinfo{journal}{Eur. Phys. J. B} \textbf{\bibinfo{volume}{32}},
  \bibinfo{pages}{429} (\bibinfo{year}{2003}{\natexlab{a}}).

\bibitem[{\citenamefont{Potthoff}(2003{\natexlab{b}})}]{pott.03.se}
\bibinfo{author}{\bibfnamefont{M.}~\bibnamefont{Potthoff}},
  \bibinfo{journal}{Eur. Phys. J. B} \textbf{\bibinfo{volume}{36}},
  \bibinfo{pages}{335} (\bibinfo{year}{2003}{\natexlab{b}}).

\bibitem[{\citenamefont{Nevidomskyy et~al.}(2008)\citenamefont{Nevidomskyy,
  S{\'en\'echal}, and Tremblay}}]{ne.se.08}
\bibinfo{author}{\bibfnamefont{A.~H.} \bibnamefont{Nevidomskyy}},
  \bibinfo{author}{\bibfnamefont{D.}~\bibnamefont{S{\'en\'echal}}},
  \bibnamefont{and} \bibinfo{author}{\bibfnamefont{A.~M.~S.}
  \bibnamefont{Tremblay}}, \bibinfo{journal}{Phys. Rev. B}
  \textbf{\bibinfo{volume}{77}}, \bibinfo{pages}{075105}
  (\bibinfo{year}{2008}).

\bibitem[{cen()}]{central}
\bibinfo{note}{To avoid confusion, we denote as ``correlated region'' the
  ``physical'' one containing interacting sites, bounded by the hoppings $V$.
  On the other hand, the ``central region'' is the one containing the clusters,
  and is bounded by $t_{bic}$. (See Fig.~\ref{laddergen})}.

\bibitem[{lea()}]{leads}
\bibinfo{note}{Obviously, the uncorrelated leads can be solved exactly without
  being partitioned into clusters}.

\bibitem[{\citenamefont{Kadanoff and Baym}(1962)}]{kad.baym}
\bibinfo{author}{\bibfnamefont{L.~P.} \bibnamefont{Kadanoff}} \bibnamefont{and}
  \bibinfo{author}{\bibfnamefont{G.}~\bibnamefont{Baym}},
  \emph{\bibinfo{title}{Quantum statistical mechanics: Green's function methods
  in equilibrium and nonequilibrium problems}}
  (\bibinfo{publisher}{Addison-Wesley}, \bibinfo{address}{Redwood City,
  Calif.}, \bibinfo{year}{1962}).

\bibitem[{\citenamefont{Schwinger}(1961)}]{schw.61}
\bibinfo{author}{\bibfnamefont{J.}~\bibnamefont{Schwinger}},
  \bibinfo{journal}{J. Math. Phys.} \textbf{\bibinfo{volume}{2}},
  \bibinfo{pages}{407} (\bibinfo{year}{1961}).

\bibitem[{\citenamefont{Keldysh}(1965)}]{keld.65}
\bibinfo{author}{\bibfnamefont{L.~V.} \bibnamefont{Keldysh}},
  \bibinfo{journal}{Sov. Phys. JETP} \textbf{\bibinfo{volume}{20}},
  \bibinfo{pages}{1018} (\bibinfo{year}{1965}).

\bibitem[{fre()}]{freq}
\bibinfo{note}{In the time representation \eqref{G} they also include
  convolutions over internal times. However, since we are considering the
  steady state, Green's functions become diagonal in the frequency
  representation}.

\bibitem[{\citenamefont{S{\'e}n{\'e}chal
  et~al.}(2000)\citenamefont{S{\'e}n{\'e}chal, Perez, and
  Pioro-Ladri{\'e}re}}]{se.pe.00}
\bibinfo{author}{\bibfnamefont{D.}~\bibnamefont{S{\'e}n{\'e}chal}},
  \bibinfo{author}{\bibfnamefont{D.}~\bibnamefont{Perez}}, \bibnamefont{and}
  \bibinfo{author}{\bibfnamefont{M.}~\bibnamefont{Pioro-Ladri{\'e}re}},
  \bibinfo{journal}{Phys. Rev. Lett.} \textbf{\bibinfo{volume}{84}},
  \bibinfo{pages}{522} (\bibinfo{year}{2000}).

\bibitem[{mno()}]{mnot}
\bibinfo{note}{Here, we use a notation to express projection of objects such as
  $G$, $T$, etc., which are matrices in lattice indices and in Keldysh space,
  onto one of the three regions $c$, $l$, or $r$. More specifically, let $m$ be
  such a matrix, then $m_{AB}$ refers to a sub-matrix of $m$ in which the left
  (right) index is restricted to region $A$ ($B$), with $A,B=c,l$, or $r$.}

\bibitem[{rea()}]{real}
\bibinfo{note}{For simplicity, we consider all hoppings to be real}.

\bibitem[{coi()}]{coincide}
\bibinfo{note}{Notice that when the central region coincides with the cluster,
  $\pcluster{G_{CC}}=G_{CC}$. In this case the solution of \eqref{vcond} is
  trivially obtained by taking the leads as auxiliary baths}.

\bibitem[{tau()}]{tau1}
\bibinfo{note}{The $\hat{\tau}_1$ in \eqref{vcond} is due to our choice of
  convention \eqref{G} for the Keldysh matrix. If one uses the form containing
  the time- and anti-time-ordered Green's functions in the diagonal, and the
  greater and lesser in the off-diagonal elements, no $\hat{\tau}_1$ is present
  in the trace}.

\bibitem[{\citenamefont{Economou}(2006)}]{economou_greens_2006}
\bibinfo{author}{\bibfnamefont{E.~N.} \bibnamefont{Economou}},
  \emph{\bibinfo{title}{Green's Functions in Quantum Physics}}
  (\bibinfo{publisher}{Springer}, \bibinfo{address}{Heidelberg},
  \bibinfo{year}{2006}).

\bibitem[{\citenamefont{{ P{\'e}rez-Merchancano} et~al.}(2008)\citenamefont{{
  P{\'e}rez-Merchancano}, Guti{\'e}rrez, and Marques}}]{pe.gu.08}
\bibinfo{author}{\bibfnamefont{S.}~\bibnamefont{{ P{\'e}rez-Merchancano}}},
  \bibinfo{author}{\bibfnamefont{H.~P.} \bibnamefont{Guti{\'e}rrez}},
  \bibnamefont{and} \bibinfo{author}{\bibfnamefont{G.~E.}
  \bibnamefont{Marques}}, \bibinfo{journal}{Microelectronics Journal}
  \textbf{\bibinfo{volume}{39}}, \bibinfo{pages}{1339 } (\bibinfo{year}{2008}),
  \bibinfo{note}{{} Papers CLACSA XIII, Colombia 2007}.

\bibitem[{\citenamefont{Chioncel et~al.}(2011)\citenamefont{Chioncel, Leonov,
  Allmaier, Beiuseanu, Arrigoni, Jurcut, and P{\"otz}}}]{ch.le.11}
\bibinfo{author}{\bibfnamefont{L.}~\bibnamefont{Chioncel}},
  \bibinfo{author}{\bibfnamefont{I.}~\bibnamefont{Leonov}},
  \bibinfo{author}{\bibfnamefont{H.}~\bibnamefont{Allmaier}},
  \bibinfo{author}{\bibfnamefont{F.}~\bibnamefont{Beiuseanu}},
  \bibinfo{author}{\bibfnamefont{E.}~\bibnamefont{Arrigoni}},
  \bibinfo{author}{\bibfnamefont{T.}~\bibnamefont{Jurcut}}, \bibnamefont{and}
  \bibinfo{author}{\bibfnamefont{W.}~\bibnamefont{P{\"otz}}},
  \bibinfo{journal}{Phys. Rev. B} \textbf{\bibinfo{volume}{83}},
  \bibinfo{pages}{035307} (\bibinfo{year}{2011}).

\bibitem[{\citenamefont{Albuquerque et~al.}(2007)\citenamefont{Albuquerque,
  Alet, Corboz, Dayal, Feiguin, Fuchs, Gamper, Gull, G{\"u}rtler, Honecker
  et~al.}}]{albuquerque_alps_2007}
\bibinfo{author}{\bibfnamefont{A.}~\bibnamefont{Albuquerque}},
  \bibinfo{author}{\bibfnamefont{F.}~\bibnamefont{Alet}},
  \bibinfo{author}{\bibfnamefont{P.}~\bibnamefont{Corboz}},
  \bibinfo{author}{\bibfnamefont{P.}~\bibnamefont{Dayal}},
  \bibinfo{author}{\bibfnamefont{A.}~\bibnamefont{Feiguin}},
  \bibinfo{author}{\bibfnamefont{S.}~\bibnamefont{Fuchs}},
  \bibinfo{author}{\bibfnamefont{L.}~\bibnamefont{Gamper}},
  \bibinfo{author}{\bibfnamefont{E.}~\bibnamefont{Gull}},
  \bibinfo{author}{\bibfnamefont{S.}~\bibnamefont{G{\"u}rtler}},
  \bibinfo{author}{\bibfnamefont{A.}~\bibnamefont{Honecker}},
  \bibnamefont{et~al.}, \bibinfo{journal}{J. Magn. Magn. Mater.}
  \textbf{\bibinfo{volume}{310}}, \bibinfo{pages}{1187} (\bibinfo{year}{2007}).

\bibitem[{imr()}]{imre}
\bibinfo{note}{For cluster DMFT, all spectral functions here $\Delta,A,\cdots$
  are matrices in cluster sites. Therefore, $\im$ is understood as the
  antihermitian part, $\re$ as the hermitian part of such matrices. The
  $v^{\pm}_n$ are column vectors in cluster sites}.

\end{thebibliography}

\end{document}